%% file: wise_nls1.tex
\documentclass[useAMS,usenatbib]{mn2e}
\usepackage{graphicx}
\usepackage{txfonts}
\usepackage{natbib}

\usepackage[english]{babel}
\usepackage{color}
\DeclareGraphicsExtensions{.pdf,.ps,.jpg}

\newcommand{\src}{SDSSJ143244.91+301435.3}
\newcommand{\sdss}{{\emph{SDSS}}}

\title[]{WISE colours and star-formation in the host galaxies of 
radio-loud narrow-line Seyfert 1}
\author[Caccianiga et al.]{A. Caccianiga$^1$,  S. Ant\'on$^{2,3}$, L. Ballo$^1$, 
L. Foschini$^1$, T. Maccacaro$^1$, R. Della~Ceca$^1$, 
\newauthor  P. Severgnini$^1$, M.J. March\~a$^4$, S. Mateos$^5$, 
E. Sani$^{6}$
\vspace{0.2cm}
\\
   $^1$INAF - Osservatorio Astronomico di Brera, via Brera 28, 
 I-20121 Milan, Italy\\
  $^2$Instituto de Astrof\'isica de Andaluc\'ia - CSIC, PO Box 3004, 18008 Granada, Spain\\
  $^3$Instituto de Astrof\'isica e Ci\^encias do Espa\c{c}o, Universidade de Lisboa
Faculdade de Ci\^encias, Campo Grande, PT1749-016 Lisboa, Portugal\\
  $^4$Department of Physics and Astronomy, University College London, Gower Street, London WC1E 6BT, UK\\
  $^5$Instituto de F\'isica de Cantabria (CSIC-Universidad de Cantabria), 39005 Santander, Spain\\
  $^6$INAF - Osservatorio Astrofisico di Arcetri, Largo E. Fermi 5, 50125 Firenze, Italy
}
  \date{}

\begin{document}

\label{firstpage}

\maketitle

\begin{abstract}
We investigate the mid-infrared properties of the largest (42 objects) 
sample of radio-loud narrow-line Seyfert 1 (RL NLS1) 
collected to date, using data from the Wide-field Infrared 
Survey Explorer (WISE). 
We analyse the mid-IR colours of these objects and 
compare 
them to what is expected from different combinations of AGN and galaxy 
templates. We find that, in general, the host galaxy emission gives an 
important
contribution to the observed mid-IR flux in particular at the longest 
wavelengths (W3, at 12\micron, and W4, at 22\micron). 
In about half of the 
sources (22 objects) we observe a very red mid-IR colour 
(W4-W3$>$2.5) that can be 
explained only using a starburst galaxy template (M82).
Using the 22\micron\ luminosities,
corrected for the AGN contribution, we have then estimated  the 
star-formation rate for 20 of these  ``red'' RL NLS1, finding values 
ranging from 10 to 500 M$_{\sun}$ y$^{-1}$. 
For the RL NLS1 showing bluer colours, instead, 
we cannot exclude the presence of a 
star-forming host galaxy although, on average, we expect a lower 
star-formation rate. 
Studying the  radio (1.4~GHz) to mid-IR (22\micron) flux 
ratios of the RL NLS1 in the sample we found that in $\sim$10 objects  
the star-forming activity could represent the 
most important component also at radio frequencies, in addition (or
in alternative) to the relativistic jet. 
We conclude that both the mid-IR and the radio emission of RL NLS1 
are a mixture of different components, including the relativistic jet, the
dusty torus and an intense star-forming activity. 

\end{abstract}


   \maketitle


\section{Introduction}
The importance of the nuclear active accretion phase (AGN) in the context of 
galaxy evolution has been fully recognized only in the last years (e.g
see \citealt{Heckman2014} for a recent review). According to the 
galaxy-AGN co-evolution scenario, at high redshifts, 
both star-forming (SF) activity and nuclear accretion must have proceeded 
simultaneously and at high rates thanks to the abundance of 
cold gas and to the frequent galaxy 
mergers that made the gas available for both processes. In a second phase,
when the AGN luminosity was high enough to expel the gas from the galaxy,
both star-formation 
and accretion rates gradually reduced eventually leading to a
``passive'' elliptical hosting a quiescent super-massive black-hole. 

At low redshift, however, galaxy interactions
and/or internal galaxy dynamics can destabilize the residual gas
and make it available for new episodes of SF and nuclear activity at
high accretion rates. \citet{Mathur2000} has proposed 
that the so-called Narrow-Line Seyfert 1 galaxies (NLS1), a peculiar
sub-class of AGN that is characterized by rapidly accreting central 
black-holes with relatively small masses (between 10$^{5}$ and 10$^{8}$ 
M$_{\sun}$ ), may represent
the low-redshift, low-luminosity analogues of the first high-z quasars,  
hosted by rejuvenated, gas-rich galaxies. 
In support to this hypothesis, it has been noted that NLS1 
are more frequently associated to host galaxies with large levels of 
star-forming 
activity than Broad Line Seyfert 1 (BLS1) galaxies (e.g. \citealt{Deo2006, 
Sani2010, Sani2012}).
In this context, NLS1 could be systems that are building up mass, evolving 
eventually into BLS1. The time-scale of this process should depend on the 
radiative efficiency of the accretion process, as discussed in 
\citet{OrbandeXivry2011}. 

Interestingly, 
NLS1 may be considered as ``young systems'' also from the radio point-of-view. 
NLS1s are typically radio-quiet (RQ) 
sources with only a small fraction that can 
be classified as radio-loud (RL, $\sim$7\%, \citealt{Komossa2006} and 
about 50 in total known to date, \citealt{Foschini2011}).
The radio properties of these RL NLS1 seem to be different from those
observed in RL BLS1 mainly because of the lack of extended radio
structures, except for very few cases (e.g. \citealt{Doi2012}). 
It has been proposed that 
at least some of the RL NLS1 can be associated to young 
radiogalaxies whose relativistic jet is still digging its way through 
the ISM of the host galaxy possibly evolving into a ``classical''
giant radiogalaxy (\citealt{Moran2000, Oshlack2001, Gallo2006, Komossa2006, 
Yuan2008, Caccianiga2014b}). 
Notably, young and compact radiogalaxies are observed also 
in the 3 radio-loud QSO discovered to date at z$\sim$6 
(\citealt{Frey2011,Frey2012}). This further increases the possible 
analogies between local NLS1 and high-z QSO.


\begin{table*}
\scriptsize 
\caption{WISE data on the sample of RL NLS1}
\label{table_wise}
\begin{tabular}{l l c  r c r  r c r  r c r  r c r}
\hline\hline
 name      & other name    &  z &     & W1   &     &       &  W2 &      &   
    &   W3 &        &     & W4 &       \\
    &     &   &      &  (3.4\micron)    &     &      &    (4.6\micron) &      &   
       &   (12 \micron) &        &     & (22\micron)  &       \\
 
           &               &    & mag & unc.    & S/N &  mag &  unc. &  S/N & mag   & unc.   &  S/N   & mag & unc. &  S/N  \\

\hline
\input{table_wise.tex}
\hline

\end{tabular}
Notes: $^1$=flagged as extended in one or more bands in the AllWISE 
catalogue (ext\_flg=3 or 5); $^2$=flagged as possible variable in W1 and W2 
band in the AllWISE catalogue (var\_flg$>$7)
\end{table*}

The possibility that NLS1 are ``young'' systems where the AGN-galaxy 
co-evolution is taking place and where, in some cases, 
a young relativistic jet is ejected and observed in the early phase of its 
evolution, is  intriguing: the study of these sources may offer a 
unique opportunity of investigating the co-evolution mechanism in both
radio-loud and radio-quiet objects at relatively 
low-redshifts where the observational constraints are less challenging with
respect to high-z QSO. 

In this paper we focus on the radio-loud fraction of NLS1 that is
still poorly studied. In particular, we want to assess if RL NLS1
are similar to RQ NLS1 for what concerns their preference in being
hosted by star-forming galaxies. To date, only few RL NLS1 have been
studied from this point of view. 
Optical imaging has revealed the 
existence of arms or circumnuclear rings (possibly the consequence of a recent
merger) and the presence of a circumnuclear starburst in at least one
RL NLS1 (1H 0323+342, \citealt{Zhou2007, Anton2008, Leon-Tavares2014}).
In a recent paper (\citealt{Caccianiga2014b}), a new
case of RL NLS1 has been presented whose 
Spectral Energy Distribution (SED) is strongly
suggestive of the presence of a quite intense star-formation 
($\sim$50 M$_{\sun}$ y$^{-1}$). The
presence of this intense SF was revealed by the mid-infrared (mid-IR) data at 
12 and 22\micron\ from the  Wide-field Infrared Survey Explorer 
(WISE) catalogue (\citealt{Wright2010}) that showed a 
clear excess with respect 
to a typical dust emission from a radio-quiet AGN or to the IR non-thermal 
spectrum coming from a jet. 
We now want to extend such a result on a large sample of RL NLS1 to
establish whether the presence of star-formation is a peculiar characteristic
of some isolated sources or, instead, is a global property of the
entire class.

The sample and the data used in this study 
are presented in Section~2. In Section~3  we discuss how 
the mid-IR colours (particularly those including the 22\micron\ band) 
of the sources cannot be satisfactorily explained by 
the AGN emission alone. In Section~4 we consider the effect of galaxy 
emission on the observed IR
colours and we infer the presence of a star-forming host in many objects
of the sample. In Section~5 we compare the 1.4~GHz to 22\micron\ 
flux ratios of the RL NLS1 with those of blazars and  infrared galaxies 
while in Section~6 we estimate the
star-formation rate (SFR) present in the sources with the
reddest W3-W4 colours. Conclusions are finally presented in Section~7.

Throughout the paper we assume a flat $\Lambda$CDM cosmology with H$_0$=71 km 
s$^{-1}$ Mpc$^{-1}$, $\Omega_{\Lambda}$=0.7 and $\Omega_{M}$=0.3. 
Spectral indices are defined assuming S$_\nu\propto\nu^{-\alpha}$.

   \begin{figure}
   \centering
    \includegraphics[width=9cm, angle=0]{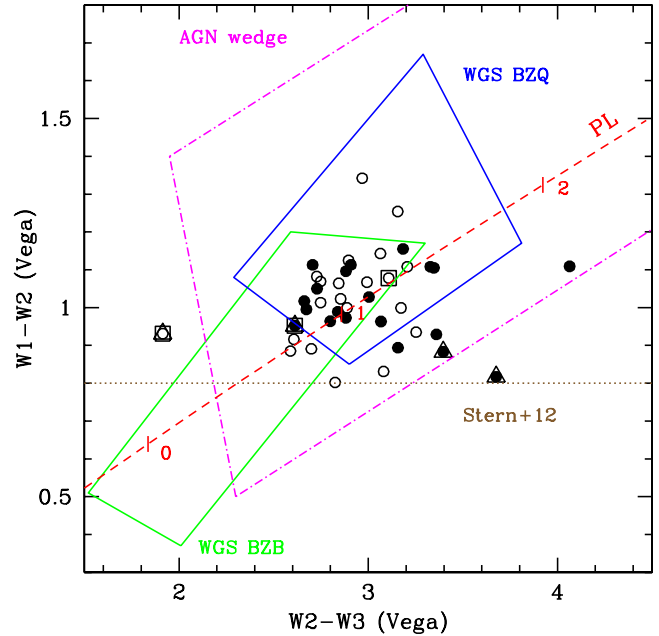}
    \caption{The WISE colours 
of the RL NLS1 in the \citet{Foschini2015} sample. Open and filled 
circles indicate
the RL NLS1 with a redshift above and below 0.3 respectively. Sources
circled by an open square 
are flagged as variable  in the 
AllWISE catalogue while sources circled by triangles 
are flagged as extended. 
Different characteristic regions used to select AGNs are also plotted:
the blazar ``WISE Gamma-ray Strip'' (WGS) for BL Lacs and FSRQ (BZB and BZQ 
respectively, solid lines, 
green and blue in the electronic version), 
as defined by \citet{Massaro2012b}, the AGN 
wedge (dot-dashed line, magenta in the electronic version) as defined 
by \citet{Mateos2012} and \citet{Mateos2013} for X-ray selected 
AGNs, the mid-IR criterion (W1-W2$>$0.8, brown dotted 
line) proposed by \citet{Stern2012}.
Finally, the red dashed line indicates the region covered by a simple power-law
spectrum (slopes of $\alpha$=0, 1 and 2 are indicated by tidemarks)
}
              \label{wise_all}
    \end{figure}

\section{The sample of RL NLS1 and WISE data}

\citet{Foschini2015} has presented an extended study of the 42 RL NLS1
with a large radio-loudness parameter\footnote{The radio-loudness
parameter used in \citet{Foschini2015} is defined as the  5~GHz to 
4400\AA\ flux density ratio.} ($>$10, see \citealt{Foschini2015} for
details) and a flat radio-spectrum (i.e. with blazar-like characteristics) 
discovered so far\footnote{We note that only 22 sources out of 42 
have a measured radio spectral index. Of these 22 objects, 
20  have a flat slope ($\alpha_R\leq$0.5) 
while 2 have $\alpha_R$ marginally above this value (0.55 and 0.58, 
respectively) 
but consistent 
with it taking into account the errors (see \citealt{Foschini2015} 
and also \citealt{Richards2015} for the recently measured spectral
index of J0953+2836).
The remaining 20 sources have no information on the 
radio spectral index and should be considered
more properly as ``flat-spectrum candidates''.}. 
Several pieces of information, from the radio
to the X-ray and gamma-ray band, have been collected and studied.
The physical properties of the inner supermassive black-hole (SMBH), like
its mass and the accretion rate, have been also discussed.
This is the largest sample of RL NLS1 collected so far and, therefore,
it represents the best starting point to study 
the incidence of star-formation in the
galaxies hosting these AGNs. It should be noted, however, that the 
type of selection of this sample, namely the requirement of a flat radio
spectrum, strongly favours the presence of a relativistically 
beamed jet emission (as in blazars), which could be 
important also in the mid-IR band. This may partly (or totally) 
hide the emission from the host galaxy. To date, however, 
only few RL NLS1 without ``blazar-like'' characteristics have been 
discovered and, therefore, the \citet{Foschini2015} sample is the 
only possibility for this type of analysis, at the moment.

The mid-IR data have been collected from the WISE all-sky
catalog (\citealt{Wright2010}) and, in particular, we have used
the last version of the catalogue, the AllWISE data release (November 2013),
which has an enhanced photometric sensitivity and accuracy compared to
the 2012 WISE All-Sky Data Release. All the NLS1 of the sample 
are detected (S/N$>$3) in the WISE survey at 3.4 and 4.6 micron 
(W1 and W2 bands, respectively), all but one are detected at
12\micron\ (W3) and 37 are detected at 22\micron\ (W4).
Data for the 42 RL NLS1 are reported in ~Table~\ref{table_wise}.

Four sources are flagged in the catalogue as possibly extended i.e.
the source profile fit has a large ($>3$)  $\chi^2$ value in one or
more bands. In one of these sources (J0324+3410/1H 0323+342)
imaging in R and B has revealed the existence of extended
structures (arms/rings) that could be the consequence of a
recent merger (\citealt{Anton2008}). In the remaining
3 sources the available optical images (from SDSS or UKST) show 
the presence of a nearby ($<$5$\arcsec$) object, possibly in interaction
with the NLS1.
Finally, 3 sources in the catalogue (J0324+3410, J0849+5108 and
J0948+0022) are flagged as variable in the W1
and W2 band and for two of them (J0849+5108 and J0948+0022)
intra-day variability has been studied by \citet{Jiang2012}.

In Fig.~\ref{wise_all} we show the distribution of the WISE colours
in the W1, W2 and W3 filters for the 41 RL NLS1 detected in all the 3 bands. 
In this figure we also show some characteristic regions recently adopted 
to select different types of AGN 
(see figure caption for details). Not surprisingly, all the objects 
fulfill the AGN definition (W1-W2$>$0.8) proposed by \citet{Stern2012} 
and all but 2 lie in the AGN wedge discussed by \citet{Mateos2012, Mateos2013}.
Most (78\%) of the RL NLS1 occupy the region where blazars 
are usually found (blue and green solid regions) while 9 sources lie outside it.
One source, in particular, shows a very blue W2-W3 colour (W2-W3=1.914),
which is significantly different from the values observed in the 
other objects. 
This source, however, is one of the 3 objects
flagged as variable (J0849+5108, one of the sources detected also
in gamma-rays by FERMI, \citealt{Foschini2015}). 
Indeed, the W1 and W2 magnitudes reported in 
\citet{Jiang2012}, based on the previous version of the WISE All-Sky
catalogue, are significantly fainter ($\Delta \sim$1.1 mag) than the
ones reported here. Therefore, the position on the W1-W2 vs W2-W3 plot of 
this objects is likely affected by the observed flux variations and, for
this reason, it is not indicative of the real spectral distribution of the 
source\footnote{
It should be considered that the AllWISE catalogue, that we are using in this
work, has been produced
by combining all the data from both the WISE cryogenic and 
post-cryogenic survey phases, the latter being carried out only in the W1 and 
W2 bands. Therefore, the periods over which the images have been collected and
combined are different for the 4 bands}.

   \begin{figure*}
   \centering
\resizebox{0.48\hsize}{!}{\includegraphics{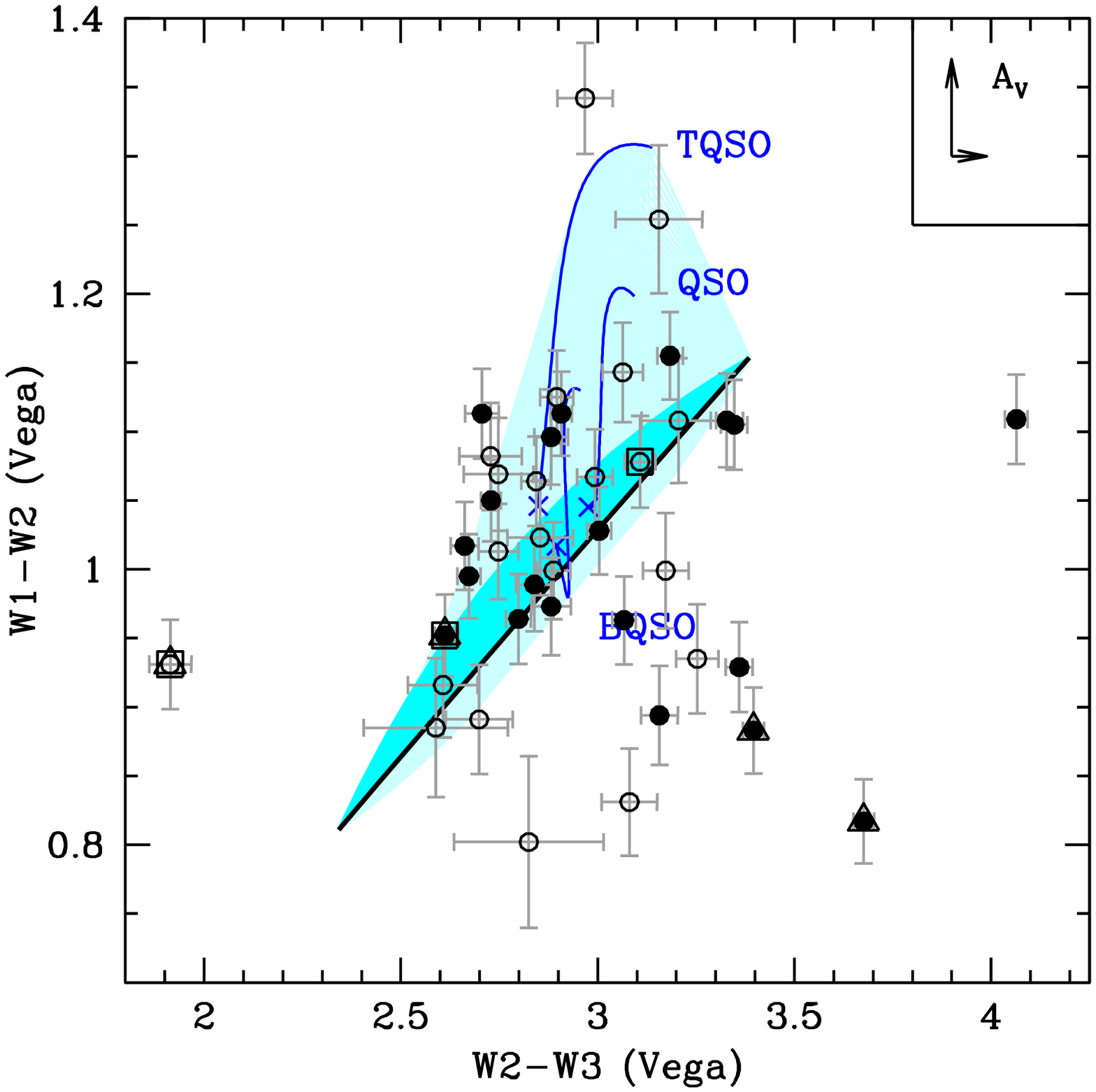}}
\resizebox{0.48\hsize}{!}{\includegraphics{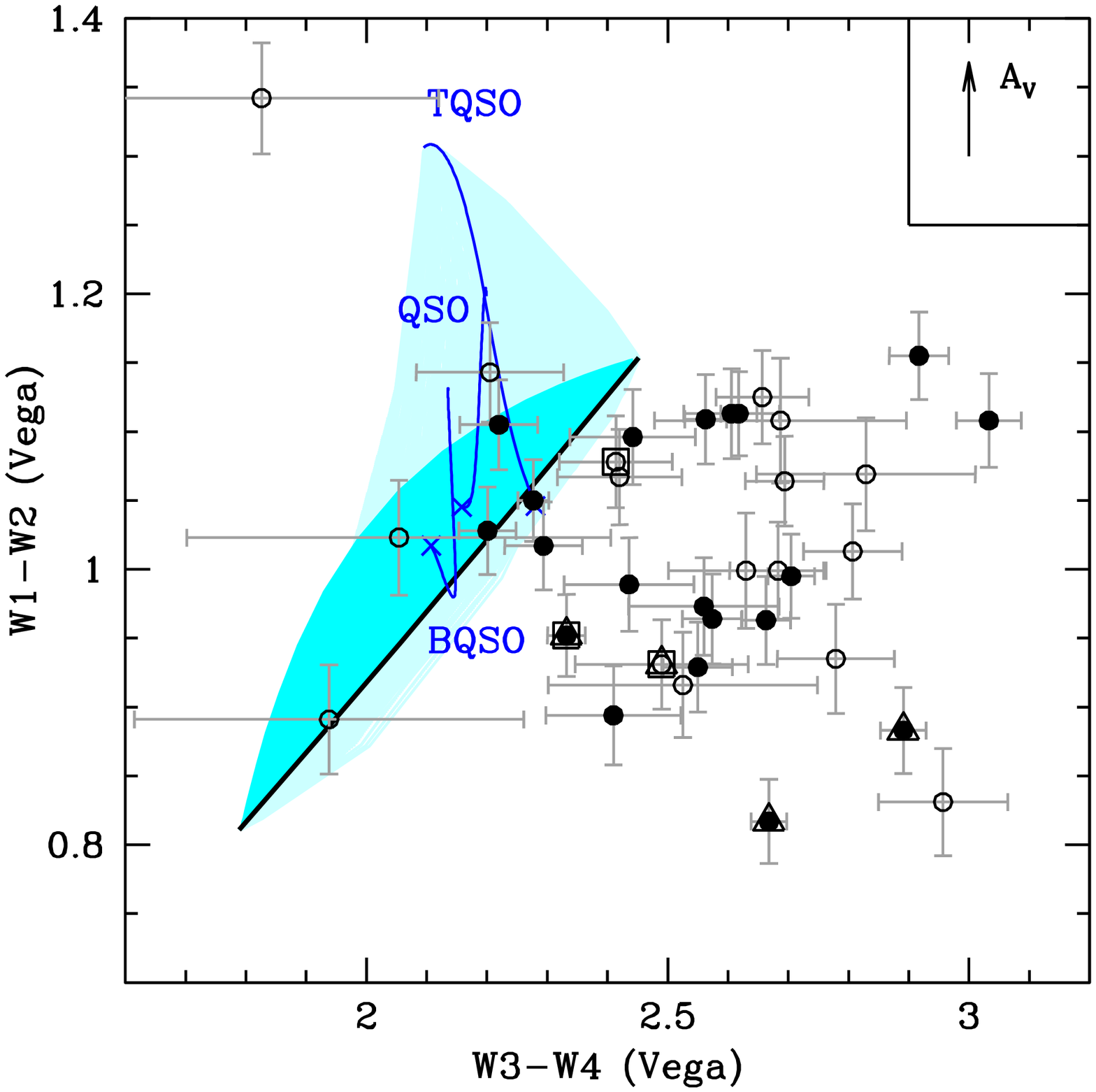}}
   \caption{The expected positions on the W1-W2 vs W2-W3 (left) and
W1-W2 vs W3-W4 (right) plot of AGNs with different shapes of IR spectrum: 
1) a simple power-law (black tick line) with slope ranging from 0.5 to 1.5;
2) a smooth broken power law, with slope before the break ranging from 
0.5 to 1 and a slope of 1.5 after the break and with the break position
varying within the WISE bands (dark blue shaded area);  
3) a RQ AGN emission described by the 3 QSO templates
from the SWIRE library (blue lines labelled as QSO, BQSO and TQSO, 
see text for a description and 
\citealt{Polletta2007} for more details) and with a redshift ranging from
0 (indicated by a cross) to 0.9 (the maximum value of the sample);
4) a combination of RQ AGN emission plus a power-law spectrum (light blue
shaded area). In particular, we show
the result of all the different combinations of slopes (from 0.5 to 1.5), 
redshift (from 0 to 0.9) and relative RQ AGN to power-law normalization.
The arrows in the upper-right corners show the expected effect of an extinction
of A$_V$=2 mag at the mean redshift of the sample. 
An A$_V$=2 mag is the maximum value expected for type~1 AGNs. 
We do not plot the arrow along the W3-W4 since the extinction has 
a negligible effect in this case.
Points as in Fig.~\ref{wise_all}}

              \label{broken_2pl}
\end{figure*}

\section{AGN emission and WISE colours}
Even if the RL NLS1 
are broadly clustered around the region expected for a power-law (PL) spectrum 
(with a spectral index $\alpha_{IR}$ ranging from $\sim$0.5 to $\sim$1.5), many 
departures from this
region are also observed: some sources occupy the region above
this line (towards redder W1-W2 colours) while others occupy a
region bluer in W1-W2 and redder in W2-W3. 

If the WISE emission of these RL NLS1 were totally dominated by the non-thermal
jet we would expect a power-law or a broken power-law  spectral shape.
In Fig.~\ref{broken_2pl} (left) we show the region on the W1-W2 vs W2-W3
plot where we expect to find 
a source whose IR spectrum is described by a broken power-law with
a slope before the break ranging from 0.5 to 1 and a slope after the break
of 1.5  and with the position of the break varying 
within the WISE bands (dark blue shaded area). 
As clear from Fig.~\ref{broken_2pl} (left), we expect that a source with
a broken power-law spectrum would fall above the PL line.  Several sources
are actually found in this region although the majority of them
seem to be much ``redder'', in terms of W1-W2 colour, than what is 
expected from a broken power-law. 

In principle a significant level of extinction could change the
position of a source with a power-law spectrum moving it away from the expected
region, towards redder values of W1-W2 and W2-W3.  
However, since these sources are, by definition, type~1 AGNs, i.e.  
objects where both the AGN continuum and the broad emission lines are visible
in the optical spectrum, the extinction is not expected to have an important
impact on the mid-IR colours (e.g. see \citealt{Mateos2015}). 
In the upper-right side of both panels of 
Fig.~\ref{broken_2pl} 
we indicate with an arrow the expected effect of an extinction
of A$_V$=2 mag at the mean redshift of the sample  ($<$z$>$=0.36). 
An A$_V$ of 2 magnitudes 
is the maximum expected for type~1 AGNs as discussed, for instance, in
\citet{Caccianiga2008}. 

It is also possible that part (or the totality) 
of the observed IR emission is produced by the radio-quiet component of the
source (i.e. the dusty torus) and not
by the non-thermal jet. Indeed, the presence of this component is expected
and it has been
revealed in many radio-loud AGN and blazars 
(e.g. \citealt{Malmrose2011, Raiteri2014, Gurkan2014, Castignani2015}). 
In Fig.~\ref{broken_2pl} (left) we plot the position of 3 radio-quiet 
type~1 QSO templates\footnote{
These templates have been derived by \citet{Polletta2007} by 
combining the \sdss\ quasar composite 
spectrum and rest-frame IR data of a sample
of optically-selected type~1 QSOs observed in the SWIRE program.
The 3 templates have the same optical spectrum but three 
different IR SEDs: a mean IR spectrum, obtained
from the average fluxes of all measurements (``QSO1''), a template with 
high IR/optical flux ratio template,
obtained from the highest $25$ per cent measurements per bin (``TQSO1''), 
and a 
low-IR emission SED obtained from the lowest
$25$ per cent measurements per bin (``BQSO1'').
} 
(from \citealt{Polletta2007}) in the range of redshift 
between 0 and 0.9 (the range of z of the sample, red solid lines).
The regions occupied by these 3 templates lie above the PL line and reach
redder W1-W2 values than the broken power-law, in particular for high 
redshift sources (z between 0.5 and 0.9). 
Notably, the source with the reddest value of W1-W2 
(=1.34, J1348+2622) is also
the object with the largest redshift in the sample (z=0.918). Only
the TQSO1 template can reach such a red W1-W2 colour that a broken
power-law, instead, does not seem able to reproduce. We consider this
as an indication that, at least in a number of objects, the 
mid-IR emission is probably mainly produced  by the radio-quiet component
(i.e. the torus) rather than the relativistic jet.

Finally, we consider also the possibility that both the dusty-torus and
the jet emissions are present in the spectrum. 
To this end, we combine the 3 QSO templates with a power-law emission, 
at different values of redshift (from 0 to 0.9), varying the relative
normalizations of the two components (light blue shaded area). 

From the analysis of Fig.~\ref{broken_2pl} (left) we conclude
that the sources lying above the PL line in the W1-W2 vs W2-W3 plot
could be explained either by a broken power-law emission, expected when
the jet emission is dominant, or by a RQ AGN (i.e. torus) emission or
by a combination of the two components. 
Only one object (J0849+5108) lies above the PL but in a 
isolated region toward blue W2-W3 values (W1-W2=0.931 and W2-W3=1.914).
As said before, this is a highly variable source and, therefore, its 
position is not indicative of the actual IR spectral shape.

In $\sim$10 cases, the objects fall well below the PL line and, therefore, 
their colours are not easily explained
by a torus/jet emission nor by a combination of the
two. The number of sources not explained by these models is even higher 
if we consider the longest wavelength magnitude (W4, 
at $\lambda$=22\micron) in the diagnostic plot i.e. if we 
plot W1-W2 vs W3-W4 (Fig.~\ref{broken_2pl}, right). As before,
we over-plot the expected regions followed by a power-law, a broken power-law 
and a RQ AGN component plus a power-law emission. 
From this figure it is clear that
only few objects can be explained by these models while the large
majority ($\sim$ 80-90\%) lie well below the PL line thus requiring a 
different (very red) component. As described in the following section,
the best candidate for the emission of this red component is
the host galaxy.

   \begin{figure}
   \centering
\includegraphics[width=9cm, angle=0]{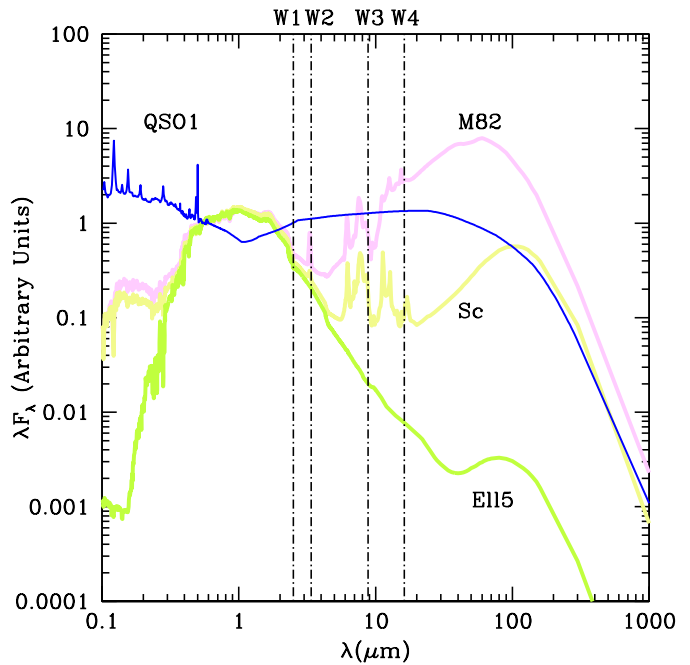}
   \caption{Templates from the SWIRE library (\citealt{Polletta2007}) used in
the analysis. 
On the top axis we indicate the centres of the four WISE bands at the
average redshift of the sample (0.36).}
              \label{show_sed}
    \end{figure}

   \begin{figure*}
   \centering
\resizebox{0.48\hsize}{!}{\includegraphics{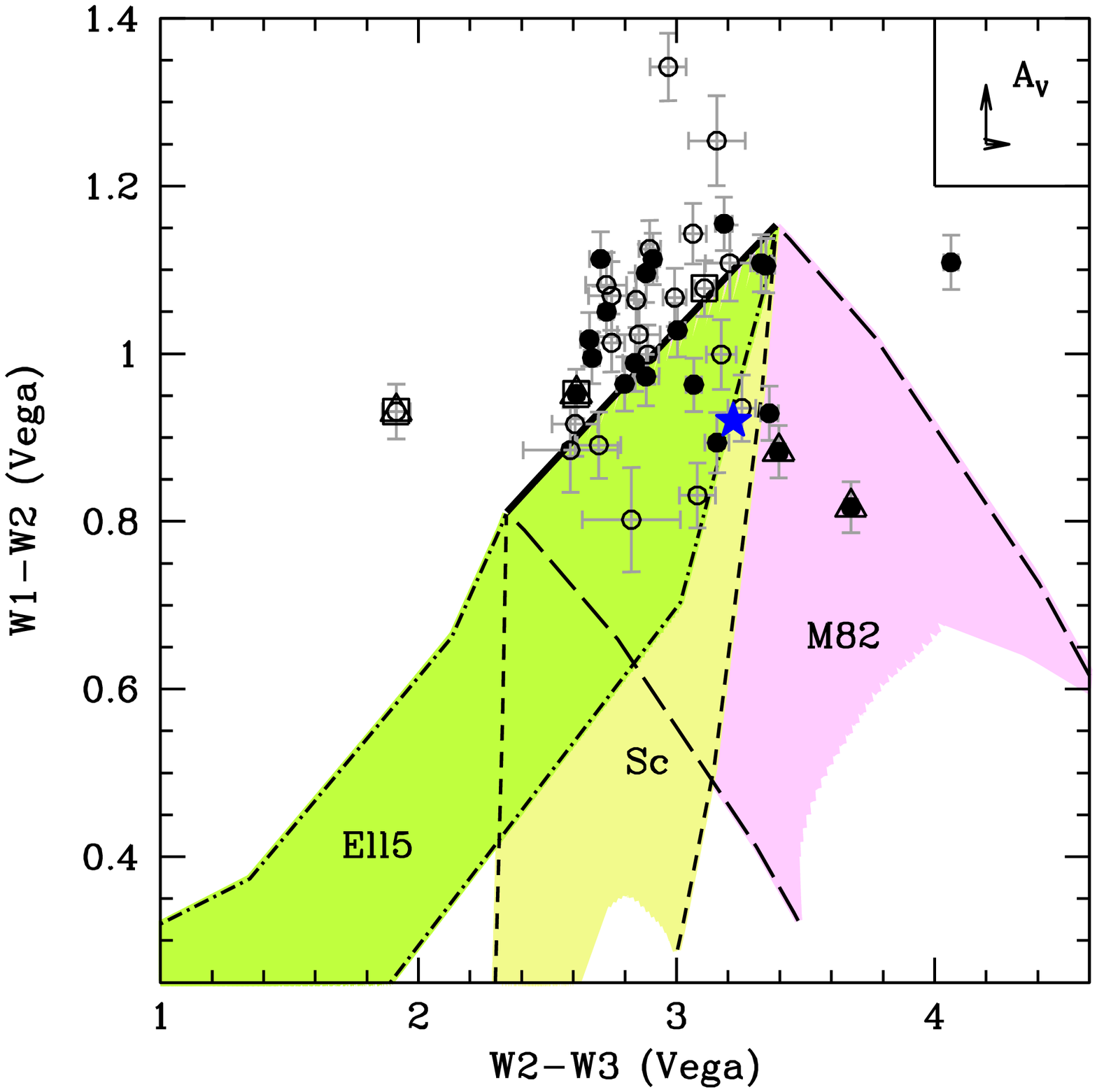}}
\resizebox{0.48\hsize}{!}{\includegraphics{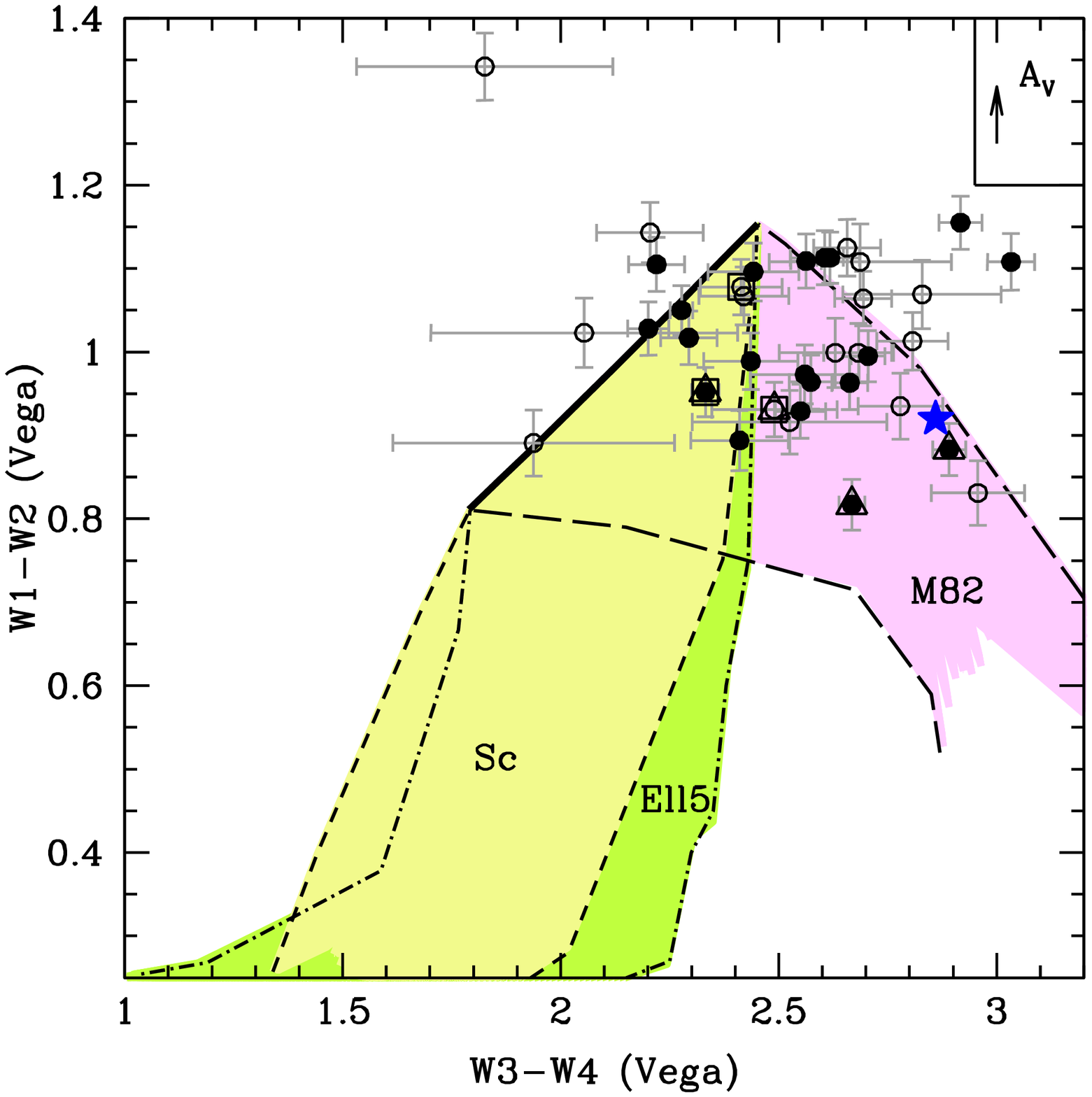}}

   \caption{
The effect of the host galaxy emission on a power-law spectrum, 
in the WISE colours plots (W1-W2 vs W2-W3, left, and vs W3-W4, right). 
The 3 regions between the long-dashed/short-dashed/point-dashed lines represent 
the expected paths of a combination of power-law plus a galaxy template 
taken from the SWIRE library (M82, Sc and Ell5 respectively, 
\citealt{Polletta2007}), varying the relative 
intensity of the two components, the power-law slopes (ranging 
from 0.5 to 1.5) and the redshift of the source ($\pm$1$\sigma$=$\pm$0.2 
from the mean value of z). Note that the limits are different from those
of Fig.~\ref{broken_2pl} in order to better show the different paths produced
by different host galaxies. 
Symbols and arrows as in Fig.~\ref{broken_2pl}. 
The blue star indicates the position of \src, a RL NLS1 studied
in detail by \citet{Caccianiga2014b} and whose SED is suggestive of the
presence of a starburst host galaxy.
}
              \label{wise_nls1_allz}
    \end{figure*}

\section{The effect of galaxy dilution on IR colours}

We consider here the effect of the combination of an AGN plus the host 
galaxy emission on the WISE colour plots. To this end, we use some
representative templates of the SWIRE library (\citealt{Polletta2007})
Specifically, we use an early-type galaxy (``Ell5'' template),
an Sc galaxy (``Sc'' template) and a starburst galaxy (``M82'' template, 
Fig.~\ref{show_sed}).
We start from a galaxy template and add a PL 
with different slopes and varying the relative 
AGN/galaxy normalization.
In Fig.~\ref{wise_nls1_allz} we show the predicted effect of the host galaxy 
dilution on a power-law emission. 
An early-type host is expected to move 
the AGN colours towards 
the left/bottom side of the plot i.e. it is expected to produce bluer
colours in both W1-W2 and W2-W3 (W3-W4) values as it gets more and more important. 
A starburst host, instead,  
is expected to make the W1-W2 values bluer than the power-law while both the W2-W3 
and the W3-W4 colours become  redder, thus 
bringing the object on the right-most side of the plot. Other starburst
templates show a similar behaviour.
This is the effect of the strong IR emission of
the starburst component that becomes more and more important in the W3 
(and W4) bands.

In Fig.~\ref{wise_nls1_allz} we also show the position of 
\src, a RL NLS1 discussed in
\citet{Caccianiga2014b} whose SED has revealed the presence
of a star-forming host galaxy with a SFR of 50 M$_{\sun}$ y$^{-1}$. 
In order to correctly reproduce the infrared SED (in particular
the W4 photometric point) of this 
object we used the M82 template by \citet{Polletta2007}. 

Fig.~\ref{wise_nls1_allz} (left) shows that the host galaxy may be
contributing significantly to the mid-IR emission in 
a number (10) of sources of the sample, i.e. those objects 
that occupy the region below the PL line. For the sources just below this
line the contribution from the host galaxy in these WISE bands is not
enough to distinguish the type of the host galaxy. 
In $\sim$4 cases, instead, the very 
red W2-W3 colours seem to favour a star-forming host galaxy. We note that
the position of \src\ on this plot does not clearly indicate a star-forming
host galaxy. Indeed, as discussed in \citet{Caccianiga2014b}, the strongest 
indication for a M82-like host galaxy comes from the W4 band (22\micron) which
shows a clear excess. 
Fig.~\ref{wise_nls1_allz} (right) shows that using the W4 filter 
we can better discriminate between
the different types of host galaxy. In particular, the 
mid-IR colours of the 22 sources with W3-W4$>$2.5 
(like \src) can be explained only by combining the AGN emission to
a starburst galaxy host.

\begin{figure*}
 \centering
\resizebox{0.48\hsize}{!}{\includegraphics{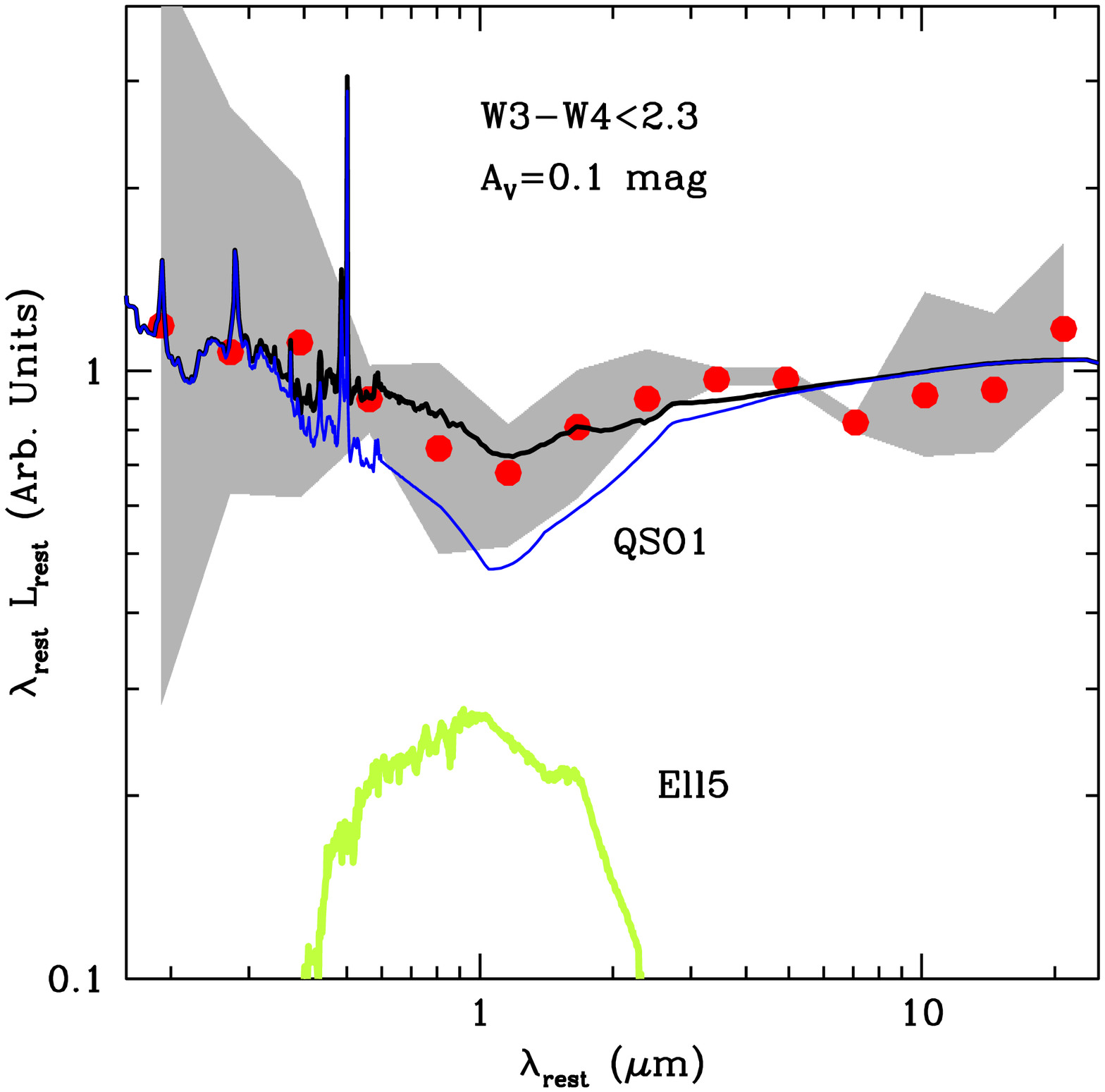}}
\resizebox{0.48\hsize}{!}{\includegraphics{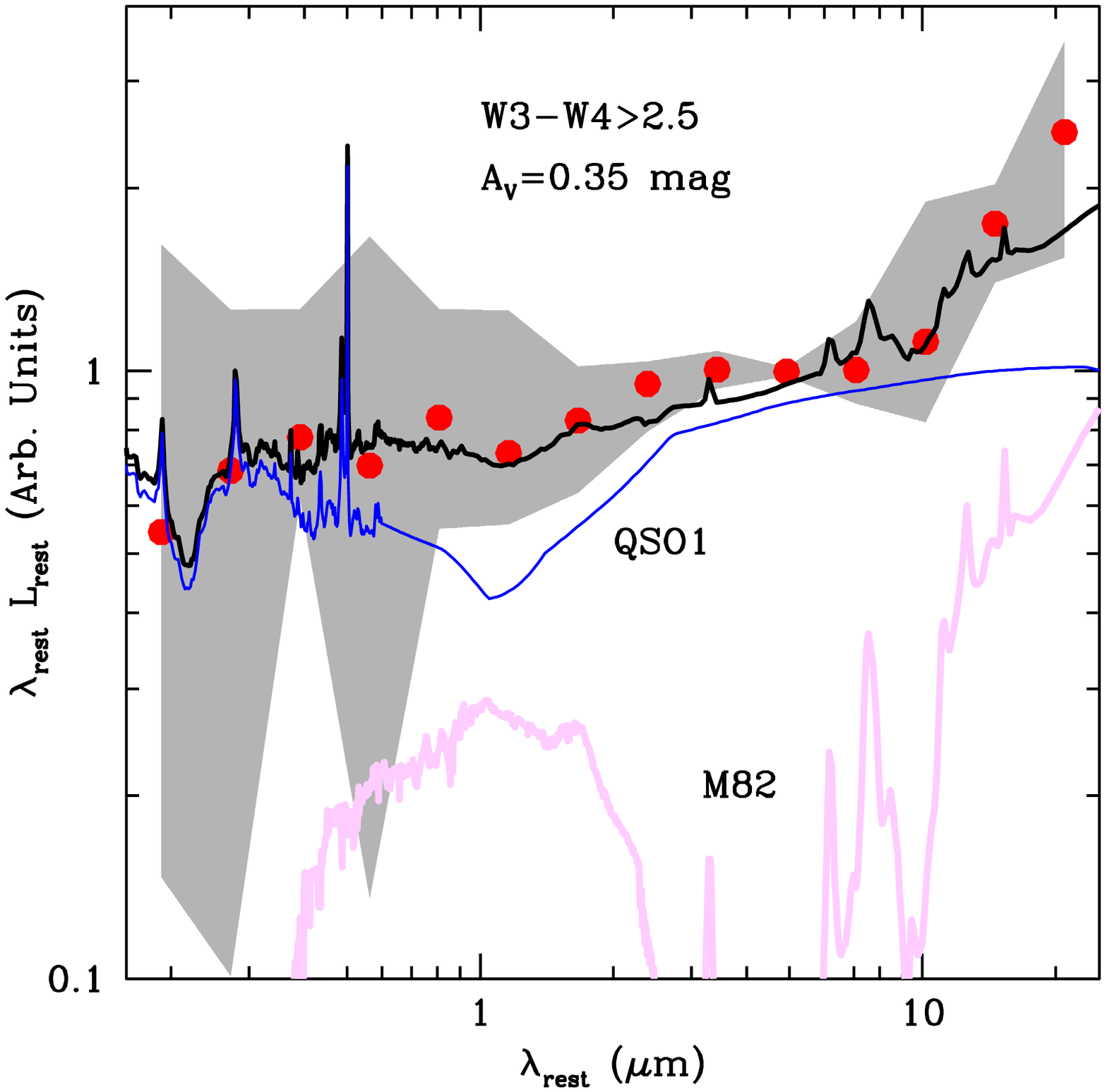}}
 \caption{Average UV to mid-IR SED of the RL NLS1
in the sample divided according to the W3$-$W4 colour. 
Red points represent the average values of Log$\lambda L_{\lambda}$ in 
each bin of Log$\lambda$ while the shaded area represents the region
including 68\% of points. 
The thick black line is a 
model composed by a reddened (the A$_V$ value is indicated in the label) 
AGN template (QSO1 from \citealt{Polletta2007}) plus a galaxy template: 
in the left panel an early-type galaxy template (Ell5) 
is used 
while, in the right panel, we adopt a starburst galaxy template (M82).}
 \label{model_avg_sed}
\end{figure*}

In order to better show the effect of the host galaxy on the observed spectrum we 
report in 
Fig.~\ref{model_avg_sed} the average UV-to-mid-IR SED of the NLS1 of the 
sample divided 
according to the value of W3-W4 colour (greater than 2.5 or lower than 2.3, in order
to keep more separated the two groups). 
The average SEDs have been created
by normalizing each rest-frame SEDs at 4\micron\ and computing the average
value of Log$\lambda$L$_\lambda$ in bins of Log$\lambda$. The grey 
area reports the region including 68\% of points. The data
used to produce these SED are taken from \citet{Foschini2015} and 
include, besides WISE photometry, data from 
Swift/UVOT observations and from existing public surveys like 
SDSS (DR9, \citealt{Ahn2012}), 2MASS (\citealt{Cutri2003}) and USNO-B
(\citealt{Monet2003}).

To correctly interpret the shape of these average SEDs we 
model them with the sum of a 
galaxy and a QSO template, varying the relative normalization, 
the amount of nuclear extinction and the host 
morphology (see \citealt{Ballo2014a} for a more detailed description of
the procedure).
We consider different combinations of extinctions and galaxy/AGN
templates and we find that templates with
a low IR emission (early-type/Sc galaxies) are able to describe
the observed average SED of the objects with bluer W3-W4 colours (left panel)
while these templates systematically fail
in reproducing the W3/W4 data points in the average SED of sources with redder 
W3-W4 (right panel). In this latter case, instead,  a starburst 
galaxy template, like M82, is more suited to reproduce 
the steep increase of the data points towards longer wavelengths.

We note that the two average SEDs show some differences also in the UV/optical
part. We can explain this difference with a different (average) extinction 
level.
While to model the average-SED of the sources with a blue W3-W4 colour
we have to apply only a small level of extinction (A$_V$=0.1 mag), in the 
case of the red W3-W4 objects, we need a higher value
of A$_V$ (0.35 mag). The explanation of this result could be
related to the enhanced presence of dust in highly star-forming galaxies with
respect to less star-forming galaxies (see e.g. \citealt{Buat2002} and 
\citealt{Dominguez2013}).

\section{Jet or star-formation?}

   \begin{figure}
   \centering
    \includegraphics[width=9cm, angle=0]{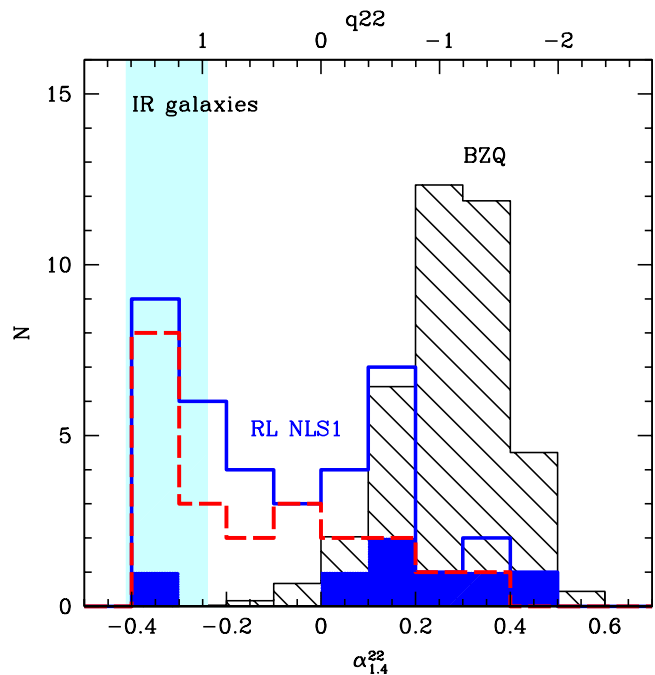}
   \caption{Distribution of the two-point spectral index between radio (1.4~GHz, rest-frame) and
mid-IR (22\micron, rest-frame) of the RL NLS1 (blue histogram) and of blazars (BZQ)
in the BZCAT (black shaded histogram). This latter histogram has been re-scaled by a factor 10
to facilitate the comparison. The flux densities have been K-corrected assuming
a slope of 0 in the radio and using the observed IR slope between 12 and 22 \micron. 
The filled blue histogram shows the RL NLS1 detected in the gamma-ray by FERMI.
The dashed red histogram includes only the red (W3-W4$>$2.5) objects.
The top axis reports the values of q22. For comparison, the typical range of q24 (defined
at 24\micron\ instead of 22\micron) observed in the
infrared galaxies is also indicated (light blue region, \citealt{Rieke2009}).
}
              \label{slope_radio_w4}

    \end{figure}

So far, we have focused our analysis on the UV to mid-IR data and, in 
particular, on the mid-IR colours. From these data we have inferred
the presence of a star-forming/starburst host galaxy in at least half 
of the RL NLS1 in the sample. 
Since an intense SF is 
expected to produce an important emission also in the radio band it is possible
that the radio luminosities observed in these sources are in part contaminated
by the SF activity. In principle, it is even possible that a high SFR can
explain the totality of the observed radio emission. This would be
an unexpected result, since the sample under investigation is composed by objects with 
large radio-loudness parameters and, in some cases, a flat radio spectrum.
These properties are usually considered as strong indications that the radio
emission is likely produced by a relativistic jet
pointing towards the observer (and, thus, relativistically beamed).

Since jet-dominated objects and star-forming galaxies have significantly different
radio-to-mid-IR flux ratios, it is interesting to compare this parameter computed for
the RL NLS1 to the values usually observed in jet-dominated
AGN (i.e. blazars), from the one hand, and to the values observed in 
star-forming galaxies, on the other hand.
 
To quantify the radio-to-mid-IR flux ratios we use two equivalent parameters. 
The two point spectral-index, defined  
between 1.4~GHz and 22\micron\  (rest-frame\footnote{For 
the K-correction we use, for each source, the slope computed between 12 and
22\micron\ (observed frame) fluxes and we assume $\alpha$=0 in the radio 
band.}) as:

\begin{equation}
\alpha_{1.4}^{22} = -\frac{Log (S_{1.4 GHz}/S_{22 \micron})}{Log (\nu_{1.4 GHz}/\nu_{22 \micron})}
\end{equation}

which is typically used in blazars studies, and the parameter q22 defined as:

\begin{equation}
q22=Log (S_{22 \micron}/S_{1.4 GHz})
\end{equation}

The radio-to-mid-IR indices for the 37 RL NLS1 in the sample detected in the W4 band
are reported in Fig.~\ref{slope_radio_w4}. For comparison, we also report the distribution
of $\alpha_{1.4}^{22}$ for a sample of 430 blazars with a quasar spectrum 
(Flat Spectrum Radio Quasar, FSRQ) extracted from the BZCAT 
(\citealt{Massaro2009}) and detected with high significance (S/N$>$10) in 
all the WISE filters. To help the comparison, this latter histogram has been re-normalized by a factor 10.
We also indicate the typical range of radio-to-mid-IR values observed in the infrared galaxies
reported by  \citet{Rieke2009}. In particular, we report the observed range of q24, a quantity
similar to q22 but defined  at 24\micron\ instead of 22\micron\ using Spitzer MIPS data. 
The average value found is 1.2-1.4,  depending on the
infrared luminosity of the source, with a dispersion of 0.24 dex.

The first consideration on Fig.~\ref{slope_radio_w4} is that the RL NLS1 span a wide interval
of  radio-to-mid-IR indices, covering the entire range between the values observed in
infrared galaxies to those observed in blazars. This is a first evidence that the radio and mid-IR 
emission of 
these sources are a mixture of different components. 
In sources with steep $\alpha_{1.4}^{22}$
indices ($>$0.2), i.e. within the range of values observed in blazars, the relativistic jet is likely
the dominant component in the radio band (while in the mid-IR different emission could be present,
including the dusty torus and/or the host galaxy).
In sources with $\alpha_{1.4}^{22}<$-0.25 (i.e. q22$>$1) the SF activity may be contributing
significantly to the observed radio emission. 
This is likely true, for instance, in the sources with red W3-W4 colours (dashed red histogram), 
that are the majority in this interval of q22 values.
The most obvious interpretation for the 
objects showing intermediate values, instead, is a combination of jet and SF (plus the dusty torus, in
the mid-IR).

In support to the hypothesis that the relativistic jet is dominant in sources with 
steep $\alpha_{1.4}^{22}$ values 
we note that 6 out of the 15 RL NLS1 with $\alpha_{1.4}^{22}>$0 (i.e. 40\%) 
are detected in gamma-rays 
by FERMI (filled histogram in Fig.~\ref{slope_radio_w4}), something that is usually 
considered as an indication for the presence 
(and the dominance) of a relativistic jet, while only one object (J1102+2239) out 22 
with $\alpha_{1.4}^{22}<$0 (5\%) is a FERMI source. The dominance of the relativistic jet in 4
of the FERMI detected RL NLS1 with q22$>$0 (J0324+3410, J0849+5108, J0948+0022 and J1505+0326) 
has been recently quantified also 
by \citet{Angelakis2014} by means of a systematic monitoring in the radio 
and millimetric bands. 

This  confirms the presence and the importance of the relativistic jet in the
sources with a positive value of $\alpha_{1.4}^{22}$ while it seems to suggest 
that other mechanisms are at work in addition (or instead of) to the relativistic jet 
in most of the remaining sources. 
It is worth noting, however, that a low value of  $\alpha_{1.4}^{22}$ (or a high value of q22) 
does not necessarily exclude the presence of a jet
as demonstrated by the detection in gamma-rays by FERMI of one of the objects 
with a high value of q22 (=1.34, J1102+2239)\footnote{Interestingly, this source 
was indicated by \citet{Foschini2015} as possible outlier in the
gamma-ray/radio correlation.}. Another example is the sources J1227+3214 that, in spite
of a large q22 value (=1.29), has an inverted radio
spectrum ($\alpha_R$=-1.04, \citealt{Foschini2015})
something that suggests a non-thermal origin of the radio emission
(or at least of a fraction of it). 
It should be considered that variability
can play an important role in determining the properties of a source at different wavelengths, 
including its position within the histogram of
Fig.~\ref{slope_radio_w4}: during high activity periods, a source may appear as
jet-dominated in many (if not all) frequencies, showing for instance a flat radio spectrum, 
a strong gamma-ray emission or a high radio-to-mid-IR flux ratio, while, during low activity periods, the 
signs for the presence of the jet can be hidden by the 
RQ AGN emission or even by the host galaxy light. 
For all these reasons, simultaneous multiwavelengths monitoring campaigns, together with 
radio follow-ups, in particular at VLBI resolution,  are fundamental to 
disentangle the different components and establish their relative importance at different
wavelengths.

   \begin{figure}
   \centering
\includegraphics[width=9cm, angle=0]{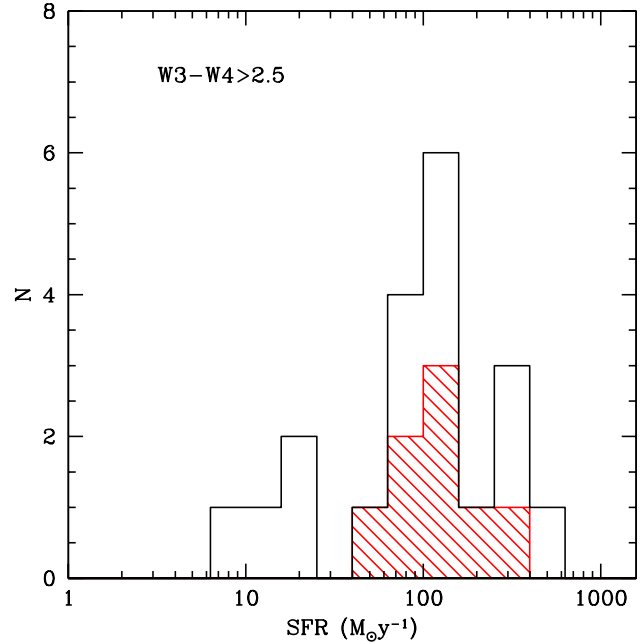}
   \caption{Star-formation rate estimated from the 22$\micron$ luminosity
corrected for the AGN contribution (see text for details). This 
estimate has been done for the objects whose WISE colours suggest
the presence of a star-forming host galaxy (W3-W4$>$2.5). Shaded 
histogram includes only the sources with a q22$>$1.
}
              \label{sfr_all_err}
    \end{figure}

\begin{table*}
\label{tab_sfr}
\begin{tabular}{r r r c r r c}
\hline\hline
 name       & Log SFR  & Log SFR unc. & q22 & W1-W2 & W3-W4 & $\alpha_R$ \\
            & (M$_{\sun}$ y$^{-1}$)  & (M$_{\sun}$ y$^{-1}$) & &  & &\\
(1) & (2) & (3) & (4) & (5) & (6) & (7) \\
\hline
\input{tab_sfr_det.tex}
\hline

\end{tabular}
\caption{Estimates of the star-formation rate for the sources with W3-W4$>$2.5 and 
with $\alpha_{1.4}^{22}<0.27$, based on the 22\micron\ luminosities corrected
for the AGN contribution (see text for details). 
Col.1: source name; col.2: logarithm of the SFR
derived from the 22\micron\ luminosity; col.3: uncertainty interval on the SFR that takes into
account only the 
errors on the estimate of the 22\micron\ luminosities of the host galaxies (see text for a description).
In addition to this uncertainty it should be considered also the one related to the SFR-L(24\micron)
relation (about $\sim$0.2 dex, \citealt{Rieke2009}); 
col. 4: logarithm of the 22\micron-to-1.4GHz flux density ratio; 
col. 5: W1-W2 colour;
col. 6: W3-W4 colour;
col. 7: radio spectral slope (S$_\nu\propto\nu^{-\alpha}$) from \citet{Foschini2015}}
\end{table*}

\section{Estimate of the star-formation rate}
In this section we want to quantify the intensity of the star-formation
at least in those sources where the mid-IR colours are suggestive of 
the presence of a starburst host. Specifically, we consider here only the 22 
sources with a red W3-W4 colour (W3-W4$>$2.5). 
To limit the contamination
due to the AGN nucleus (either the dusty torus or the jet component)
we analyse here the longest wavelength WISE band (W4 at 22\micron) since,
according to the results previously discussed, this is the band where
the star-forming nature of the host galaxy seems to emerge more clearly.

We first compute the 22$\micron$ luminosities ($\lambda L_\lambda$).
We derive the 22$\micron$ flux densities from the W4 magnitudes\footnote{
For this conversion we assume an IR spectral slope of 1. The differences in the
derived flux densities when assuming significantly
different slopes, from -1 to 2, are below 0.7\%} and applied   
a K-correction using the spectral slope computed between 12 and
22\micron\ (observed frame) fluxes.  
The computed 22$\micron$ luminosities of the sources 
range between 10$^{10}$ to 
10$^{12}$ L$_{\sun}$ with a large fraction (76\%) of objects with 
L$_{22\micron}>$10$^{11}$ L$_{\sun}$. 

In order to derive an estimate of the SFR in 
these sources we can use the relation presented in \citet{Rieke2009} between
SFR and L$_{24\micron}$ (their eq.10 and eq.11).
However, since the L$_{22\micron}$ luminosity is likely contaminated by the
AGN, in particular the dusty torus component, we have to apply a correction. 
To estimate the importance of the AGN emission
at 22\micron\ we proceed as follows.

{\it Jet contribution}. We estimate the contribution from the jet starting from the
radio flux density measured at 1.4~GHz, assuming the
average value of the $\alpha_{1.4}^{22}$ observed in the
blazars (FSRQ) of the BZCAT catalogue ($\alpha_{1.4}^{22}$=0.27,
see Fig.~\ref{slope_radio_w4}) 
and subtract it from the observed flux density at 22$\micron$ (rest-frame).
For the 2 objects that have $\alpha_{1.4}^{22}>$0.27 the estimate
of 22$\micron$ due to the galaxy is not possible and we exclude them
from the analysis. For the sources with radio-to-mid-IR flux
ratios similar to those observed in IR galaxies ($\alpha_{1.4}^{22}<$-0.25),
the correction for the jet emission is negligible ($<$1\%) and has no
impact on the final results.  As discussed in the previous section, in these
sources even the radio emission could be mainly (or totally) attributed to the
star-formation activity.

{\it Torus contribution}. We estimate the contribution from the torus using 
the flux densities at 4.6$\micron$ (rest frame\footnote{For the 
K-correction we use here, for each source, 
the observed slope computed between 3.4 and 4.6 \micron.}) and assuming that
this is entirely produced by the AGN (jet and torus). 
This is a reasonable assumption 
since, as discussed in the previous sections, in the W1 and W2 filters
we are probably  preferentially observing the AGN emission (see also
the average SEDs of Fig.~\ref{model_avg_sed}). This is also
consistent with what has been recently found by \citet{Mateos2015} studying
the mid-IR properties of a sample of X-ray selected AGN.
If also
the host galaxy is contributing to the 4.6\micron\ emission then the
final estimate of the 22\micron\ emission from the host galaxy, and hence
of the SFR, would be even higher.
As for the 22\micron\ flux density, also the 4.6\micron\ flux density
has been corrected for the jet contribution, using the average 1.4~GHz-to-4.6\micron\
spectral index observed in the FSRQ of the BZCAT ($<\alpha_{1.4}^{4.6}>$=0.43). 
Using the resulting flux density and the slope  between 4.6 and 22\micron\ 
measured on the QSO1 template from SWIRE ($\alpha$=1.1), 
we can then estimate the contribution of the torus at 22$\micron$ and
subtract also this component from the observed flux density. 

We find that the average contribution from the torus at 22\micron\ is 
more important ($\sim$40\%) than that from the jet ($\sim$7\%). 
In order to take into account the uncertainties related to the
initial assumption on the $\alpha_{1.4}^{22}$ of the jet and
the IR slope of the torus, we have repeated these computations with
different input values: $\alpha_{1.4}^{22}$=0.15 and 0.39 (i.e. 
the 1$\sigma$ interval observed in the FSRQ of the BZCAT) 
and $\alpha_{IR}$=1.0-1.1, for the infrared slope of the torus,
corresponding to the extreme values measured in the 3 QSO1 template
of SWIRE (TQSO1, BQSO1 and QSO1). We
take the extreme values obtained in all these combinations as the 
uncertainty interval of the 22\micron\ luminosity from the 
host galaxy.
The computed 22\micron\ luminosities of the host galaxies
range from 10$^{10}$ to 5$\times$10$^{11}$L$_{\sun}$.

We then use the 22$\micron$ luminosity corrected for the
AGN contribution to estimate the SFR using the relation presented in
\citet{Rieke2009}, their eq.10 and eq.11, and assuming L$_{24\micron} (MIPS) \sim$L$_{22\micron}$.
The results are reported in Tab.~\ref{tab_sfr} and plotted in 
Fig.~\ref{sfr_all_err}. 
The SFR of these 20 objects 
are above $\sim$10 $M_{\sun}$ y$^{-1}$ and, in several cases, larger than 
100 $M_{\sun}$ y$^{-1}$, similarly to what is usually observed in 
Luminous Infrared Galaxies (LIRG) or even in Ultraluminous Infrared Galaxies (ULIRG,
\citealt{Sanders1996}). This result confirms and quantifies what we have inferred from
the analysis of the mid-IR colours, i.e. the presence of an intense
star-forming activity in at least half of the RL NLS1 in sample.

We finally note that a SFR estimate for the RL NLS1 with W3-W4$<$2.5 
would not be 
reliable since the evidence of a significant contribution of a star-forming host galaxy 
to the emission at 22\micron\ is not clear. Running the analysis described
above on the blue objects, we find that, on average, only a small fraction ($\sim$8\%)
of the observed 22$\micron$ luminosities can be due to the host galaxy the remaining
fraction being produced by the torus (56\%) and by the jet (36\%). 
The estimate of the 22\micron\ luminosity due to the host galaxy (and therefore
the SFR quantification) depends too critically on the starting assumptions
on the $\alpha_{1.4}^{22}$ of the jet and on the IR spectral shape of the dusty torus
to give reliable results. The only stable quantities that we can derive are the 
upper limits on the SFR based on the total observed 22\micron\ luminosities.
These values do not exclude the presence of an intense SFR 
($>$ 10 $M_{\sun}$ y$^{-1}$) also in the objects with W3-W4$<$2.5.
From these upper limits alone it is not possible to understand whether the 
host galaxies of these AGN are intrinsically
different from those observed in ``red'' RL NLS1 or, rather, their SF
activity is simply hidden by a particularly strong nuclear light. 
We have, however, other pieces of information that can help to distinguish
between these two hypotheses.

In Fig.~\ref{model_avg_sed} we have presented the average SEDs of red (W3-W4$>$2.5) 
and blue (W3-W4$<$2.3) objects, that  clearly show different shapes at $\lambda>$10\micron. 
While the average SED of the red objects requires the presence of an
M82-like template to be correctly reproduced, the average SED of the blue 
objects can be modelled using an Sc or Ell5 template. An M82 template 
would over-predict the emission at $\lambda>$10\micron. We note that 
the normalization of the host galaxy template is quite well constrained in the 
spectral region around $\sim$1\micron\ (rest-frame). 
This suggests that the blue objects
are hosted, on average, by different types of host galaxy. 

In order to assess whether there are differences also in the 
luminosites of the AGN (either the radio-quiet or the jet component) between
red and blue objects we can compare the average luminosities at 4\micron,
where the torus emission is expected to dominate, and
in the radio band. At 4\micron\ the average values
are only marginally different, with the blue objects being
about 30\% more luminous than the red ones. 
Even in the
radio band we do not have evidence that the jets in the blue objects are more luminous
than the jets present in the red sources. 
Rather, the average 1.4~GHz power of blue objects is
a factor $\sim$1.6 fainter than in  red objects.
The average 22\micron\ luminosity, instead, is significantly different
bewteeen the two groups of sources, where the red objects are almost a factor
2 more luminous that the blue sources.

Overall, these numbers seem to indicate that the lack of a clear detection of the host galaxy
in the blue objects is not due to a more luminous radio jet and/or AGN (radio-quiet) 
but, rather, to a less luminous  22\micron\ 
emission, i.e. a less important star-forming component at work in these sources.
As explained above, the upper limits on the SFR are in any case large so we cannot
exclude the presence of an intense SF also in these sources but, on average,
their activity, if present, seems to be less intense than the one observed in the
host galaxies of the red RL NLS1. 
This picture needs a confirmation through an accurate follow-up (in particular
of the blue RL NLS1) in the mid-IR or in the far-IR spectral region, 
where the maximum output of the SF activity is expected.

\section{Conclusions}

We have analysed the mid-IR colours of a sample of 42 RL NLS1 using WISE data
with the aim of quantifying, for the first time, the star-forming activity
in the galaxies hosting this class of AGNs. The sources cover the redshift 
interval from 0.06 to 0.92. 
Thirty-seven objects are detected in all the 4 WISE bands. 
In order to understand the origin of the mid-IR emission in these sources
we have compared their positions on different 
diagnostic plots with those expected from 
a combination of AGN and host galaxy templates.  
The results can be summarized as follows:

\begin{itemize}

\item In general, the RL NLS1 of the sample occupy a region in the 
W1-W2 vs W2-W3 plot (commonly used in the literature) that is typical
of emission line AGN, both radio-quiet and radio-loud.

\item WISE colours can be reproduced by a combination of AGN and 
host galaxy emission, the relative importance of the
two components  depending strongly on the band under consideration.  
While the emission from 3.4$\micron$ up to 12$\micron$ (observed
frame) of most (3/4) of the sources can be explained by the
AGN alone (either the relativistic
jet, the dusty torus or a combination of the two), at 22\micron\
the emission usually requires the (red) contribution from the host galaxy.

\item Twenty-two sources show very red 
W3-W4 colours ($>$2.5) that can be explained only by assuming the 
presence of a young and ``active'' host galaxy, i.e. a starburst
galaxy similar to M82.  In the remaining cases, the intensity of the 
AGN emission in the mid-IR bands does not allow to unambiguously assess the
type of host galaxy.

\item The 22$\micron$ 
luminosities of the NLS1 detected in W4 are typically above 
10$^{11}$ L$_{\sun}$. We have used these luminosities,  corrected  for the 
AGN contribution, to estimate the SFR for 20 out 
of the 22 sources with W3-W4$>$2.5. 
The computed values of SFR range between 10 and 500 M$_{\sun}$ y$^{-1}$.
These values are similar to those observed in LIRG or even in ULIRG 
(\citealt{Sanders1996}). For the sources with W3-W4$<$2.5 
the presence of a star-forming host galaxy cannot be excluded
although the SFR is expected to be, on average, lower than in red sources.

\item Although the sample studied here has been designed to include 
NLS1 with a relativistic jet (possibly pointing towards Earth), 
by means of a selection
of sources with a large radio-loudness parameter (radio-loudness parameter 
$>$10) and, preferentially, 
a flat radio spectrum ($\alpha_R<$0.5, see 
\citealt{Foschini2015} for details), our analysis shows that in 
the mid-IR the jet is not necessarily the dominant
component, even in sources that have been detected in gamma-rays by 
FERMI. In particular, we have estimated that, on average, at 22\micron\ only
$\sim$20\% of the luminosity of the sources with W3-W4$>$2.5 comes from the jet, 
the remaining part being emitted by the dusty torus and by the host 
galaxy.

\item Even at radio frequencies (1.4 GHz) the emission of a number ($\sim$10) of
sources in the sample, i.e. those with high mid-IR-to-radio luminosity
ratios, is not necessarily due to the jet but
it is likely produced, in part or entirely, by the SF activity. 

\end{itemize}

In conclusion, our analysis has shown that RL NLS1 are often associated to
``active'' host galaxies, with SFR in the typical range of
LIRG/ULIRG sources.
Therefore, from this point of view, RL NLS1 are more similar to RQ NLS1
than to radio-loud AGNs with broader emission lines, usually
hosted by ``passive'' elliptical galaxies. These results support the
idea that NLS1, both RQ and RL, are systems in the early phase of their
evolution, when the host galaxy is experiencing a high level of star-forming
activity and, at the same time, the central super-massive BH is rapidly accreting and
building up mass. Studying the properties of these sources 
and their differences with respect to the other radio-loud AGNs will help us to
understand the role of relativistic jets in the AGN/galaxy evolution and to unveil 
any possible form of feedback (either positive or negative). 
To this end, simultaneous follow-ups, both in the mid-IR and in the radio, are 
mandatory in order to accurately disentangle the different components 
at work in these active nuclei and to study their possible interplay.

\section*{Acknowledgments}
We thank the referee for useful comments that improved the paper. 
This publication makes use of data products from the Wide-field Infrared Survey Explorer, 
which is a joint project of the University of California, Los Angeles, and the Jet Propulsion 
Laboratory/California Institute of Technology, funded by the National Aeronautics and Space 
Administration.
Part of this work was supported by the COST Action MP0905 ``Black Holes in a Violent Universe'' 
and by the European Commission Seventh Framework
Programme (FP7/2007-2013) under grant agreement no. 267251
Astronomy Fellowships in Italy (AstroFIt).
The authors acknowledge financial support from the Italian Ministry of Education, 
Universities and Research (PRIN2010-2011, grant n.
2010NHBSBE). 
Support from the Italian Space Agency is acknowledged by LB (contract ASI INAF NuSTAR I/037/12/0).
SM acknowledges funding from the Spanish Ministry of Economy and Competitiveness under 
grant AYA2012-31447, which is partly funded by the FEDER programme. SM acknowledges financial 
support from the ARCHES project (7th Framework of the European Union, no. 313146).

\bibliographystyle{mn2e}
\bibliography{/home/guincho/caccia/cartella_comune/NLS1_WISE}

\end{document}

%% file: table_wise.tex
J0100-0200 & FBQSJ0100-0200                &  0.227 & 12.859 &  0.023 &   46.5 & 11.754 &  0.023 &   46.7 &          8.407 &          0.024 &  44.7 &          6.187 &          0.060 &  18.0 \\
J0134-4258 & PMNJ0134-4258                 &  0.237 & 12.157 &  0.024 &   46.1 & 11.129 &  0.021 &   52.4 &          8.125 &          0.022 &  49.1 &          5.924 &          0.042 &  25.9 \\
J0324+3410 & 1H0323+342 03             $^1$ $^2$ &  0.061 & 10.743 &  0.022 &   49.1 &  9.791 &  0.020 &   54.7 &          7.179 &          0.016 &  67.4 &          4.847 &          0.027 &  40.1 
     \\
J0706+3901 & FBQSJ0706+3901                &  0.086 & 12.888 &  0.024 &   45.6 & 11.959 &  0.022 &   48.9 &          8.599 &          0.026 &  41.0 &          6.049 &          0.051 &  21.4 \\
J0713+3820 & FBQSJ0713+3820                &  0.123 & 10.040 &  0.022 &   49.3 &  8.990 &  0.020 &   53.5 &          6.261 &          0.015 &  74.8 &          3.984 &          0.021 &  51.7 \\
J0744+5149 & NVSSJ074402+514917            &  0.460 & 13.397 &  0.024 &   44.4 & 12.384 &  0.025 &   43.7 &          9.636 &          0.044 &  24.7 &          6.829 &          0.069 &  15.6 \\
J0804+3853 & SDSSJ080409.23+385348.8       &  0.211 & 11.745 &  0.023 &   46.4 & 10.750 &  0.020 &   54.3 &          8.077 &          0.022 &  48.8 &          5.372 &          0.032 &  33.5 \\
J0814+5609 & SDSSJ081432.11+560956.6       &  0.509 & 14.221 &  0.027 &   40.4 & 13.139 &  0.028 &   38.9 &         10.411 &          0.074 &  14.6 & n.d. & - & - \\
J0849+5108 & SDSSJ084957.97+510829.0   $^1$ $^2$ &  0.584 & 12.887 &  0.024 &   44.7 & 11.956 &  0.022 &   48.8 &         10.042 &          0.049 &  22.0 &          7.552 &          0.135 &   8.0 
     \\
J0902+0443 & SDSSJ090227.16+044309.5       &  0.532 & 13.927 &  0.026 &   41.0 & 13.096 &  0.029 &   37.7 &         10.015 &          0.064 &  16.9 &          7.058 &          0.086 &  12.7 \\
J0937+3615 & SDSSJ093703.02+361537.1       &  0.179 & 12.165 &  0.024 &   45.7 & 11.202 &  0.021 &   51.5 &          8.135 &          0.021 &  50.8 &          5.472 &          0.035 &  31.1 \\
J0945+1915 & SDSSJ094529.23+191548.8       &  0.284 & 12.042 &  0.024 &   45.9 & 11.078 &  0.022 &   50.1 &          8.279 &          0.023 &  47.4 &          5.705 &          0.044 &  25.0 \\
J0948+0022 & SDSSJ094857.31+002225.4     $^2$ &  0.585 & 13.282 &  0.024 &   44.5 & 12.204 &  0.023 &   46.5 &          9.096 &          0.032 &  33.4 &          6.682 &          0.088 &  12.3 
  \\
J0953+2836 & SDSSJ095317.09+283601.5       &  0.658 & 14.809 &  0.032 &   33.9 & 13.924 &  0.039 &   27.6 &         11.335 &          0.179 &   6.1 & n.d. & - & - \\
J1031+4234 & SDSSJ103123.73+423439.3       &  0.376 & 14.323 &  0.029 &   37.8 & 13.300 &  0.030 &   35.8 &         10.446 &          0.078 &  14.0 &          8.392 &          0.343 &   3.2 \\
J1037+0036 & SDSSJ103727.45+003635.6       &  0.595 & 15.108 &  0.037 &   29.5 & 13.854 &  0.039 &   27.8 &         10.698 &          0.103 &  10.6 & n.d. & - & - \\
J1038+4227 & SDSSJ103859.58+422742.2       &  0.220 & 12.549 &  0.024 &   46.0 & 11.532 &  0.021 &   50.7 &          8.870 &          0.028 &  39.1 &          6.576 &          0.058 &  18.8 \\
J1047+4725 & SDSSJ104732.68+472532.0       &  0.798 & 14.749 &  0.031 &   35.2 & 13.641 &  0.033 &   33.3 &         10.435 &          0.089 &  12.2 &          7.748 &          0.189 &   5.7 \\
J1048+2222 & SDSSJ104816.58+222239.0       &  0.330 & 13.499 &  0.025 &   44.3 & 12.356 &  0.026 &   42.4 &          9.292 &          0.044 &  24.5 &          7.087 &          0.114 &   9.5 \\
J1102+2239 & SDSSJ110223.38+223920.7       &  0.453 & 13.167 &  0.024 &   45.7 & 12.042 &  0.024 &   44.7 &          9.146 &          0.034 &  31.5 &          6.489 &          0.069 &  15.7 \\
J1110+3653 & SDSSJ111005.03+365336.3       &  0.630 & 16.038 &  0.058 &   18.6 & 15.230 &  0.090 &   12.1 & n.d. & - & - & n.d. & - & - \\
J1138+3653 & SDSSJ113824.54+365327.1       &  0.356 & 14.055 &  0.026 &   42.1 & 13.139 &  0.028 &   39.1 &         10.532 &          0.084 &  13.0 &          8.007 &          0.207 &   5.2 \\
J1146+3236 & SDSSJ114654.28+323652.3       &  0.465 & 14.099 &  0.027 &   39.8 & 13.208 &  0.029 &   38.0 &         10.509 &          0.080 &  13.6 &          8.571 &          0.313 &   3.5 \\
J1159+2838 & SDSSJ115917.32+283814.5       &  0.210 & 13.451 &  0.025 &   42.6 & 12.343 &  0.023 &   46.3 &          9.015 &          0.034 &  32.2 &          5.982 &          0.042 &  25.8 \\
J1227+3214 & SDSSJ122749.14+321458.9       &  0.137 & 11.441 &  0.023 &   48.2 & 10.328 &  0.020 &   53.7 &          7.420 &          0.018 &  60.8 &          4.802 &          0.024 &  44.8 \\
J1238+3942 & SDSSJ123852.12+394227.8       &  0.623 & 15.327 &  0.037 &   29.7 & 14.525 &  0.050 &   21.9 &         11.700 &          0.183 &   5.9 & n.d. & - & - \\
J1246+0238 & SDSSJ124634.65+023809.0       &  0.363 & 14.163 &  0.029 &   37.4 & 13.094 &  0.029 &   37.2 &         10.346 &          0.084 &  12.9 &          7.517 &          0.161 &   6.7 \\
J1333+4141 & SDSSJ133345.47+414127.7       &  0.225 & 13.102 &  0.023 &   47.2 & 11.947 &  0.022 &   49.5 &          8.763 &          0.024 &  44.6 &          5.846 &          0.043 &  25.1 \\
J1346+3121 & SDSSJ134634.97+312133.7       &  0.246 & 13.744 &  0.025 &   43.1 & 12.850 &  0.026 &   41.8 &          9.693 &          0.039 &  27.6 &          7.283 &          0.105 &  10.3 \\
J1348+2622 & SDSSJ134834.28+262205.9       &  0.918 & 14.710 &  0.029 &   37.0 & 13.368 &  0.028 &   38.2 &         10.400 &          0.064 &  16.9 &          8.574 &          0.287 &   3.8 \\
J1358+2658 & SDSSJ135845.38+265808.5       &  0.331 & 13.133 &  0.024 &   44.7 & 12.069 &  0.022 &   48.3 &          9.225 &          0.030 &  36.2 &          6.531 &          0.058 &  18.6 \\
J1421+2824 & SDSSJ142114.05+282452.8       &  0.538 & 13.674 &  0.025 &   43.0 & 12.607 &  0.024 &   45.8 &          9.614 &          0.038 &  28.7 &          7.194 &          0.096 &  11.3 \\
J1505+0326 & SDSSJ150506.47+032630.8       &  0.409 & 14.028 &  0.027 &   40.1 & 13.093 &  0.029 &   37.8 &          9.840 &          0.045 &  24.4 &          7.061 &          0.086 &  12.7 \\
J1548+3511 & SDSSJ154817.92+351128.0       &  0.479 & 13.770 &  0.026 &   42.1 & 12.771 &  0.024 &   44.8 &          9.883 &          0.035 &  31.4 &          7.200 &          0.072 &  15.1 \\
J1612+4219 & SDSSJ161259.83+421940.3       &  0.234 & 13.308 &  0.024 &   45.8 & 12.199 &  0.022 &   48.9 &          8.135 &          0.018 &  62.0 &          5.572 &          0.030 &  36.3 \\
J1629+4007 & SDSSJ162901.30+400759.9       &  0.272 & 13.290 &  0.024 &   45.6 & 12.177 &  0.022 &   49.1 &          9.471 &          0.037 &  29.4 &          6.865 &          0.070 &  15.6 \\
J1633+4718 & SDSSJ163323.58+471858.9   $^1$   &  0.116 & 12.135 &  0.023 &   46.6 & 11.318 &  0.020 &   53.2 &          7.642 &          0.017 &  64.0 &          4.974 &          0.024 &  44.8 
  \\
J1634+4809 & SDSSJ163401.94+480940.2       &  0.495 & 14.883 &  0.028 &   39.0 & 13.884 &  0.031 &   35.1 &         10.711 &          0.049 &  22.1 &          8.081 &          0.119 &   9.2 \\
J1644+2619 & SDSSJ164442.53+261913.2       &  0.145 & 13.283 &  0.024 &   44.7 & 12.294 &  0.024 &   44.4 &          9.455 &          0.040 &  27.1 &          7.019 &          0.100 &  10.8 \\
J1709+2348 & SDSSJ170907.80+234837.7       &  0.254 & 13.758 &  0.026 &   42.5 & 12.785 &  0.024 &   44.4 &          9.903 &          0.044 &  24.6 &          7.343 &          0.117 &   9.2 \\
J2007-4434 & PKS 2004-447                  &  0.240 & 13.445 &  0.025 &   43.7 & 12.349 &  0.024 &   44.8 &          9.467 &          0.035 &  31.1 &          7.025 &          0.098 &  11.0 \\
J2021-2235 & IRAS 20181-2244           $^1$   &  0.185 & 11.880 &  0.023 &   48.0 & 10.997 &  0.021 &   51.6 &          7.601 &          0.017 &  63.5 &          4.710 &          0.034 &  32.2 
  \\

%% file: tab_sfr_det.tex
J0706+3901 &  0.95  & [  0.92 ,   0.98 ] &  0.82 &  0.93 &  2.55 & -     \\
J0744+5149 &  2.50  & [  2.43 ,   2.56 ] &  0.47 &  1.01 &  2.81 & -     \\
J0804+3853 &  1.95  & [  1.94 ,   1.96 ] &  1.48 &  0.99 &  2.70 & -     \\
J0902+0443 &  2.70  & [  1.70 ,   3.70 ] & -0.65 &  0.83 &  2.96 & 0.07  $ \pm $  0.01 \\
J0937+3615 &  1.92  & [  1.91 ,   1.93 ] &  1.31 &  0.96 &  2.66 & -     \\
J0945+1915 &  2.14  & [  2.06 ,   2.21 ] &  0.60 &  0.96 &  2.57 & -     \\
J1102+2239 &  2.48  & [  2.47 ,   2.49 ] &  1.34 &  1.13 &  2.66 & -     \\
J1138+3653 &  1.16  & [  0.16 ,   2.16 ] & -0.15 &  0.92 &  2.52 & 0.5  $ \pm $  0.09 \\
J1159+2838 &  2.11  & [  2.10 ,   2.11 ] &  1.39 &  1.11 &  3.03 & -     \\
J1227+3214 &  1.73  & [  1.71 ,   1.74 ] &  1.29 &  1.11 &  2.62 & -1.04  $ \pm $  0.07 \\
J1246+0238 &  1.83  & [  0.83 ,   2.83 ] & -0.36 &  1.07 &  2.83 & 0.55  $ \pm $  0.06 \\
J1333+4141 &  2.17  & [  2.16 ,   2.17 ] &  1.39 &  1.16 &  2.92 & -     \\
J1358+2658 &  2.11  & [  2.09 ,   2.12 ] &  1.30 &  1.06 &  2.69 & -     \\
J1548+3511 &  2.18  & [  1.18 ,   3.18 ] & -0.77 &  1.00 &  2.68 & 0.26  $ \pm $  0.01 \\
J1612+4219 &  2.35  & [  2.35 ,   2.36 ] &  1.32 &  1.11 &  2.56 & -     \\
J1629+4007 &  1.39  & [  1.06 ,   1.72 ] &  0.29 &  1.11 &  2.61 & -0.68  $ \pm $  0.02 \\
J1633+4718 &  1.83  & [  1.74 ,   1.92 ] &  0.23 &  0.82 &  2.67 & 0.42  $ \pm $  0.01 \\
J1634+4809 &  2.07  & [  1.93 ,   2.20 ] &  0.15 &  1.00 &  2.63 & -     \\
J1709+2348 &  1.36  & [  1.33 ,   1.39 ] &  0.96 &  0.97 &  2.56 & -     \\
J2021$-$2235 &  2.48  & [  2.46 ,   2.50 ] &  0.81 &  0.88 &  2.89 & 0.5  $ \pm $  0.07 \\

%% file: wise_nls1.bbl
\begin{thebibliography}{44}
\expandafter\ifx\csname natexlab\endcsname\relax\def\natexlab#1{#1}\fi

\bibitem[{Ahn {et~al}\mbox{.}(2012)Ahn, Alexandroff, {Allende Prieto},
  Anderson, Anderton, Andrews, Aubourg, Bailey, Balbinot, Barnes, Bautista,
  Beers, Beifiori, Berlind, Bhardwaj, Bizyaev, Blake, Blanton, Blomqvist,
  Bochanski, Bolton, Borde, Bovy, Brandt, Brinkmann, Brown, Brownstein, Bundy,
  Busca, Carithers, Carnero, Carr, Casetti-Dinescu, Chen, Chiappini, Comparat,
  Connolly, Crepp, Cristiani, Croft, Cuesta, da~Costa, Davenport, Dawson,
  de~Putter, {De Lee}, Delubac, Dhital, Ealet, Ebelke, Edmondson, Eisenstein,
  Escoffier, Esposito, Evans, Fan, {Femen\'{\i}a Castell\'{a}}, {Fern\'{a}ndez
  Alvar}, Ferreira, {Filiz Ak}, Finley, Fleming, Font-Ribera, Frinchaboy,
  Garc\'{\i}a-Hern\'{a}ndez, P\'{e}rez, Ge, G\'{e}nova-Santos, Gillespie,
  Girardi, {Gonz\'{a}lez Hern\'{a}ndez}, Grebel, Gunn, Guo, Haggard, Hamilton,
  Harris, Hawley, Hearty, Ho, Hogg, Holtzman, Honscheid, Huehnerhoff, Ivans,
  Ivezi\'{c}, Jacobson, Jiang, Johansson, Johnson, Kauffmann, Kirkby,
  Kirkpatrick, Klaene, Knapp, Kneib, {Le Goff}, Leauthaud, Lee, Lee, Long,
  Loomis, Lucatello, Lundgren, Lupton, Ma, Ma, MacDonald, Mack, Mahadevan,
  Maia, Majewski, Makler, Malanushenko, Malanushenko, Manchado, Mandelbaum,
  Manera, Maraston, Margala, Martell, McBride, McGreer, McMahon, M\'{e}nard,
  Meszaros, Miralda-Escud\'{e}, Montero-Dorta, Montesano, Morrison, Muna, Munn,
  Murayama, Myers, Neto, Nguyen, Nichol, Nidever, Noterdaeme, Nuza, Ogando,
  Olmstead, Oravetz, Owen, Padmanabhan, Palanque-Delabrouille, Pan, Parejko,
  Parihar, P\^{a}ris, Pattarakijwanich, Pepper, Percival, P\'{e}rez-Fournon,
  P\'{e}rez-R\`{a}fols, Petitjean, Pforr, Pieri, Pinsonneault, {Porto de
  Mello}, Prada, Price-Whelan, Raddick, Rebolo, Rich, Richards, Robin,
  Rocha-Pinto, Rockosi, Roe, Ross, Ross, Rossi, Rubi\~{n}o Martin, Samushia,
  {Sanchez Almeida}, S\'{a}nchez, Santiago, Sayres, Schlegel, Schlesinger,
  Schmidt, Schneider, Schultheis, Schwope, Sc\'{o}ccola, Seljak, Sheldon, Shen,
  Shu, Simmerer, Simmons, Skibba, Skrutskie, Slosar, Sobreira, Sobeck, Stassun,
  Steele, Steinmetz, Strauss, Streblyanska, Suzuki, Swanson, Tal, Thakar,
  Thomas, Thompson, Tinker, Tojeiro, Tremonti, {Vargas Maga\~{n}a}, Verde,
  Viel, Vikas, Vogt, Wake, Wang, Weaver, Weinberg, Weiner, West, White, Wilson,
  Wisniewski, Wood-Vasey, Yanny, Y\`{e}che, York, Zamora, Zasowski, Zehavi,
  Zhao, Zheng, Zhu, \& Zinn}]{Ahn2012}
Ahn C.~P. {et~al.}, 2012, \apjs, 203, 21

\bibitem[{Angelakis {et~al}\mbox{.}(2015)Angelakis, Fuhrmann, Marchili,
  Foschini, Myserlis, Karamanavis, Komossa, Blinov, Krichbaum, Sievers,
  Ungerechts, \& Zensus}]{Angelakis2014}
Angelakis E. {et~al.}, 2015, \aap, 575, 55

\bibitem[{Ant\'{o}n, Browne \& March\~{a}(2008)Ant\'{o}n, Browne, \&
  March\~{a}}]{Anton2008}
Ant\'{o}n S., Browne I. W.~A., March\~{a} M.~J., 2008, \aap, 490, 583

\bibitem[{Ballo {et~al}\mbox{.}(2014)Ballo, Severgnini, {Della Ceca},
  Caccianiga, Vignali, Carrera, Corral, \& Mateos}]{Ballo2014a}
Ballo L., Severgnini P., {Della Ceca} R., Caccianiga A., Vignali C., Carrera
  F.~J., Corral A., Mateos S., 2014, \mnras, 444, 2580

\bibitem[{Buat {et~al}\mbox{.}(2002)Buat, Boselli, Gavazzi, \&
  Bonfanti}]{Buat2002}
Buat V., Boselli A., Gavazzi G., Bonfanti C., 2002, \aap, 383, 801

\bibitem[{Caccianiga {et~al}\mbox{.}(2014)Caccianiga, Ant\'{o}n, Ballo,
  Dallacasa, {Della Ceca}, Fanali, Foschini, Hamilton, Kraus, Maccacaro, Mack,
  March\~{a}, Paulino-Afonso, Sani, \& Severgnini}]{Caccianiga2014b}
Caccianiga A. {et~al.}, 2014, \mnras, 441, 172

\bibitem[{Caccianiga {et~al}\mbox{.}(2008)Caccianiga, Severgnini, {Della Ceca},
  Maccacaro, Cocchia, Barcons, Carrera, Matute, McMahon, Page, Pietsch,
  Sbarufatti, Schwope, Tedds, \& Watson}]{Caccianiga2008}
Caccianiga A. {et~al.}, 2008, \aap, 477, 735

\bibitem[{Castignani \& {De Zotti}(2015)}]{Castignani2015}
Castignani G., {De Zotti} G., 2015, \aap, 573, 125

\bibitem[{Cutri {et~al}\mbox{.}(2003)Cutri, Skrutskie, van Dyk, Beichman,
  Carpenter, Chester, Cambresy, Evans, Fowler, Gizis, Howard, Huchra, Jarrett,
  Kopan, Kirkpatrick, Light, Marsh, McCallon, Schneider, Stiening, Sykes,
  Weinberg, Wheaton, Wheelock, \& Zacarias}]{Cutri2003}
Cutri R.~M. {et~al.}, 2003, "The IRSA 2MASS All-Sky Point Source Catalog",
  NASA/IPAC

\bibitem[{Deo, Crenshaw \& Kraemer(2006)Deo, Crenshaw, \& Kraemer}]{Deo2006}
Deo R.~P., Crenshaw D.~M., Kraemer S.~B., 2006, \aj, 132, 321

\bibitem[{Doi {et~al}\mbox{.}(2012)Doi, Nagira, Kawakatu, Kino, Nagai, \&
  Asada}]{Doi2012}
Doi A., Nagira H., Kawakatu N., Kino M., Nagai H., Asada K., 2012, \apj, 760,
  41

\bibitem[{Dom\'{\i}nguez {et~al}\mbox{.}(2013)Dom\'{\i}nguez, Siana, Henry,
  Scarlata, Bedregal, Malkan, Atek, Ross, Colbert, Teplitz, Rafelski, McCarthy,
  Bunker, Hathi, Dressler, Martin, \& Masters}]{Dominguez2013}
Dom\'{\i}nguez A. {et~al.}, 2013, \apj, 763, 145

\bibitem[{Foschini(2011)}]{Foschini2011}
Foschini L., 2011, in: Narrow-Line Seyfert 1 Galaxies and Their Place in the
  Universe, eds L. Foschini, M. Colpi, L. Gallo, D. Grupe, S. Komossa, K.
  Leighly, \& S. Mathur. Proceedings of Science (Trieste, Italy), vol. NLS1,
  id. 24

\bibitem[{Foschini {et~al}\mbox{.}(2015)Foschini, Berton, Caccianiga, Ciroi,
  Cracco, Peterson, Angelakis, Braito, Gallo, Grupe, J\"{a}rvel\"{a}, Kaufmann,
  Komossa, \& Kovalev}]{Foschini2015}
Foschini L. {et~al.}, 2015, A\&A, 575, A13

\bibitem[{Frey {et~al}\mbox{.}(2012)Frey, Gurvits, Paragi, \&
  Gabanyi}]{Frey2012}
Frey S., Gurvits L.~I., Paragi Z., Gabanyi K., 2012, in: "Resolving The Sky -
  Radio Interferometry: Past, Present and Future". Proceedings of Science
  (Trieste, Italy), vol. RTS2012, id. 41

\bibitem[{Frey {et~al}\mbox{.}(2011)Frey, Paragi, Gurvits, Gab\'{a}nyi, \&
  Cseh}]{Frey2011}
Frey S., Paragi Z., Gurvits L.~I., Gab\'{a}nyi K.~E., Cseh D., 2011, \aap, 531,
  L5

\bibitem[{Gallo {et~al}\mbox{.}(2006)Gallo, Edwards, Ferrero, Kataoka, Lewis,
  Ellingsen, Misanovic, Welsh, Whiting, Boller, Brinkmann, Greenhill, \&
  Oshlack}]{Gallo2006}
Gallo L.~C. {et~al.}, 2006, \mnras, 370, 245

\bibitem[{Gurkan, Hardcastle \& Jarvis(2014)Gurkan, Hardcastle, \&
  Jarvis}]{Gurkan2014}
Gurkan G., Hardcastle M.~J., Jarvis M.~J., 2014, \mnras, 438, 1149

\bibitem[{Heckman \& Best(2014)}]{Heckman2014}
Heckman T.~M., Best P.~N., 2014, \araa, 52, 75

\bibitem[{Jiang {et~al}\mbox{.}(2012)Jiang, Zhou, Ho, Yuan, Wang, Dong, Jiang,
  Ji, \& Tian}]{Jiang2012}
Jiang N. {et~al.}, 2012, \apj, 759, L31

\bibitem[{Komossa {et~al}\mbox{.}(2006)Komossa, Voges, Xu, Mathur, Adorf,
  Lemson, Duschl, \& Grupe}]{Komossa2006}
Komossa S., Voges W., Xu D., Mathur S., Adorf H.-M., Lemson G., Duschl W.~J.,
  Grupe D., 2006, \aj, 132, 531

\bibitem[{Le\'{o}n-Tavares {et~al}\mbox{.}(2014)Le\'{o}n-Tavares, Kotilainen,
  Chavushyan, A\~{n}orve, Puerari, Cruz-Gonz\'{a}lez, Pati\~{n}o \'{A}lvarez,
  Ant\'{o}n, Carrami\~{n}ana, Carrasco, Guichard, Karhunen,
  Olgu\'{\i}n-Iglesias, Sanghvi, \& Valdes}]{Leon-Tavares2014}
Le\'{o}n-Tavares J. {et~al.}, 2014, \apj, 795, 58

\bibitem[{Malmrose {et~al}\mbox{.}(2011)Malmrose, Marscher, Jorstad, Nikutta,
  \& Elitzur}]{Malmrose2011}
Malmrose M.~P., Marscher A.~P., Jorstad S.~G., Nikutta R., Elitzur M., 2011,
  \mnras, 732, 116

\bibitem[{Massaro {et~al}\mbox{.}(2009)Massaro, Giommi, Leto, Marchegiani,
  Maselli, Perri, Piranomonte, \& Sclavi}]{Massaro2009}
Massaro E., Giommi P., Leto C., Marchegiani P., Maselli A., Perri M.,
  Piranomonte S., Sclavi S., 2009, \aap, 495, 691

\bibitem[{Massaro {et~al}\mbox{.}(2012)Massaro, D'Abrusco, Tosti, Ajello,
  Gasparrini, Grindlay, \& Smith}]{Massaro2012b}
Massaro F., D'Abrusco R., Tosti G., Ajello M., Gasparrini D., Grindlay J.~E.,
  Smith H.~A., 2012, \apj, 750, 138

\bibitem[{Mateos {et~al}\mbox{.}(2013)Mateos, Alonso-Herrero, Carrera, Blain,
  Severgnini, Caccianiga, \& Ruiz}]{Mateos2013}
Mateos S., Alonso-Herrero A., Carrera F., Blain A., Severgnini P., Caccianiga
  A., Ruiz A., 2013, \mnras, 434, 941

\bibitem[{Mateos {et~al}\mbox{.}(2012)Mateos, Alonso-Herrero, Carrera, Blain,
  Watson, Barcons, Braito, Severgnini, Donley, \& Stern}]{Mateos2012}
Mateos S. {et~al.}, 2012, \mnras, 426, 3271

\bibitem[{Mateos {et~al}\mbox{.}(2015)Mateos, Carrera, Rovilos, Hern, Barcons,
  Blain, Caccianiga, Ceca, \& Severgnini}]{Mateos2015}
Mateos S. {et~al.}, 2015, \mnras, 449, 1422

\bibitem[{Mathur(2000)}]{Mathur2000}
Mathur S., 2000, \mnras, 314, L17

\bibitem[{Monet {et~al}\mbox{.}(2003)Monet, Levine, Canzian, Ables, Bird, Dahn,
  Guetter, Harris, Henden, Leggett, Levison, Luginbuhl, Martini, Monet, Munn,
  Pier, Rhodes, Riepe, Sell, Stone, Vrba, Walker, Westerhout, Brucato, Reid,
  Schoening, Hartley, Read, \& Tritton}]{Monet2003}
Monet D.~G. {et~al.}, 2003, \aj, 125, 984

\bibitem[{Moran(2000)}]{Moran2000}
Moran E.~C., 2000, New Astronomy Reviews, 44, 527

\bibitem[{{Orban de Xivry} {et~al}\mbox{.}(2011){Orban de Xivry}, Davies,
  Schartmann, Komossa, Marconi, Hicks, Engel, \& Tacconi}]{OrbandeXivry2011}
{Orban de Xivry} G., Davies R., Schartmann M., Komossa S., Marconi A., Hicks
  E., Engel H., Tacconi L., 2011, \mnras, 417, 2721

\bibitem[{Oshlack, Webster \& Whiting(2001)Oshlack, Webster, \&
  Whiting}]{Oshlack2001}
Oshlack A. Y. K.~N., Webster R.~L., Whiting M.~T., 2001, \apj, 558, 578

\bibitem[{Polletta {et~al}\mbox{.}(2007)Polletta, Tajer, Maraschi, Trinchieri,
  Lonsdale, Chiappetti, Andreon, Pierre, {Le Fevre}, Zamorani, Maccagni,
  Garcet, Surdej, Franceschini, Alloin, Shupe, Surace, Fang, Rowan‐Robinson,
  Smith, \& Tresse}]{Polletta2007}
Polletta M. {et~al.}, 2007, \apj, 663, 81

\bibitem[{Raiteri {et~al}\mbox{.}(2014)Raiteri, Villata, Carnerero,
  Acosta-Pulido, Larionov, D'Ammando, Ar\'{e}valo, Arkharov, Bueno, {Di Paola},
  Efimova, Gonz\'{a}lez-Morales, Gorshanov, Grinon-Marin, L\'{a}zaro,
  Manilla-Robles, Yabar, Gim\'{e}nez, \& Velasco}]{Raiteri2014}
Raiteri C.~M. {et~al.}, 2014, \mnras, 442, 629

\bibitem[{Richards \& Lister(2015)}]{Richards2015}
Richards J.~L., Lister M.~L., 2015, \apj, 800, L8

\bibitem[{Rieke {et~al}\mbox{.}(2009)Rieke, Alonso-Herrero, Weiner,
  P\'{e}rez-Gonz\'{a}lez, Blaylock, Donley, \& Marcillac}]{Rieke2009}
Rieke G.~H., Alonso-Herrero A., Weiner B.~J., P\'{e}rez-Gonz\'{a}lez P.~G.,
  Blaylock M., Donley J.~L., Marcillac D., 2009, \apj, 692, 556

\bibitem[{Sanders \& Mirabel(1996)}]{Sanders1996}
Sanders D.~B., Mirabel I.~F., 1996, \araa, 34, 749

\bibitem[{Sani {et~al}\mbox{.}(2012)Sani, Davies, Sternberg, Graci\'{a}-Carpio,
  Hicks, Krips, Tacconi, Genzel, Vollmer, Schinnerer, Garc\'{\i}a-Burillo,
  Usero, \& {Orban de Xivry}}]{Sani2012}
Sani E. {et~al.}, 2012, \mnras, 424, 1963

\bibitem[{Sani {et~al}\mbox{.}(2010)Sani, Lutz, Risaliti, Netzer, Gallo,
  Trakhtenbrot, Sturm, \& Boller}]{Sani2010}
Sani E., Lutz D., Risaliti G., Netzer H., Gallo L.~C., Trakhtenbrot B., Sturm
  E., Boller T., 2010, \mnras, 403, 1246

\bibitem[{Stern {et~al}\mbox{.}(2012)Stern, Assef, Benford, Blain, Cutri, Dey,
  Eisenhardt, Griffith, Jarrett, Lake, Masci, Petty, Stanford, Tsai, Wright,
  Yan, Harrison, \& Madsen}]{Stern2012}
Stern D. {et~al.}, 2012, \apj, 753, 30

\bibitem[{Wright {et~al}\mbox{.}(2010)Wright, Eisenhardt, Mainzer, Ressler,
  Cutri, Jarrett, Kirkpatrick, Padgett, McMillan, Skrutskie, Stanford, Cohen,
  Walker, Mather, Leisawitz, Gautier, McLean, Benford, Lonsdale, Blain, Mendez,
  Irace, Duval, Liu, Royer, Heinrichsen, Howard, Shannon, Kendall, Walsh,
  Larsen, Cardon, Schick, Schwalm, Abid, Fabinsky, Naes, \& Tsai}]{Wright2010}
Wright E.~L. {et~al.}, 2010, \aj, 140, 1868

\bibitem[{Yuan {et~al}\mbox{.}(2008)Yuan, Zhou, Komossa, Dong, Wang, Lu, \&
  Bai}]{Yuan2008}
Yuan W., Zhou H.~Y., Komossa S., Dong X.~B., Wang T.~G., Lu H.~L., Bai J.~M.,
  2008, \apj, 685, 801

\bibitem[{Zhou {et~al}\mbox{.}(2007)Zhou, Wang, Yuan, Shan, Komossa, Lu, Liu,
  Xu, Bai, \& Jiang}]{Zhou2007}
Zhou H. {et~al.}, 2007, \apjl, 658, L13

\end{thebibliography}
